\documentclass[12pt,preprint]{aastex}

\usepackage{epsfig}
\usepackage{lscape,graphicx,rotating,amsmath,color}

\newcommand{\Swift}{\textit{Swift}}

\newcommand{\rategrbz}{{\tt RATE GRB-$z$}}
\newcommand{\rate}{{\tt RATE}}
\begin{document}

\title{Rapid, Machine-Learned Resource Allocation: Application to High-redshift GRB Follow-up}

\author{A.~N.~Morgan\altaffilmark{1,*}, James~Long\altaffilmark{2}, Joseph~W.~Richards\altaffilmark{1,2}, Tamara~Broderick\altaffilmark{2}, Nathaniel~R.~Butler\altaffilmark{1,3}, and Joshua~S.~Bloom\altaffilmark{1}}

\altaffiltext{1}{Department of Astronomy, University of
  California, Berkeley, CA 94720-3411, USA. }
\altaffiltext{2}{Department of Statistics, University of California, Berkeley, CA 94720-3860, USA}
\altaffiltext{3}{Department of Physics, Arizona State University, Tempe, AZ 85287, USA}
\altaffiltext{*}{\url{amorgan@astro.berkeley.edu}}

\slugcomment{Accepted to \apj}

\shorttitle{Rapid ML Resource Allocation for High-z GRBs}
\shortauthors{Morgan \textit{et al.}}

\begin{abstract}

As the number of observed Gamma-Ray Bursts (GRBs) continues to grow, follow-up resources need to be used more efficiently in order to maximize science output from limited telescope time. As such, it is becoming increasingly important to  rapidly identify bursts of interest as soon as possible after the event, before the afterglows fade beyond detectability.  Studying the most distant (highest redshift) events, for instance, remains a primary goal for many in the field.  Here we present our Random forest Automated Triage Estimator for GRB redshifts (\rategrbz{}) for rapid identification of high-redshift candidates using early-time metrics from the three telescopes onboard \textit{Swift}. While the basic \rate{} methodology is generalizable to a number of resource allocation problems, here we demonstrate its utility for telescope-constrained follow-up efforts with the primary goal to identify and study high-$z$ GRBs.  For each new GRB, \rategrbz{}  provides a recommendation---based on the available telescope time---of whether the event warrants additional follow-up resources.  We train  \rategrbz{}   using a set consisting of 135 \Swift{} bursts with known redshifts, only 18 of which are $z > 4$.  Cross-validated performance metrics on this training data suggest that $\sim$56\% of high-$z$ bursts can be captured from following up the top 20\% of the ranked candidates, and $\sim$84\% of high-$z$ bursts are identified after following up the top $\sim$40\% of candidates.  We further use the method to rank $ 200+$ \Swift{}  bursts with unknown redshifts according to their likelihood of being high-$z$.

\end{abstract}


\section{Introduction}
\label{sec:intro}

\begin{figure}[t!]
  \centerline{\plotone{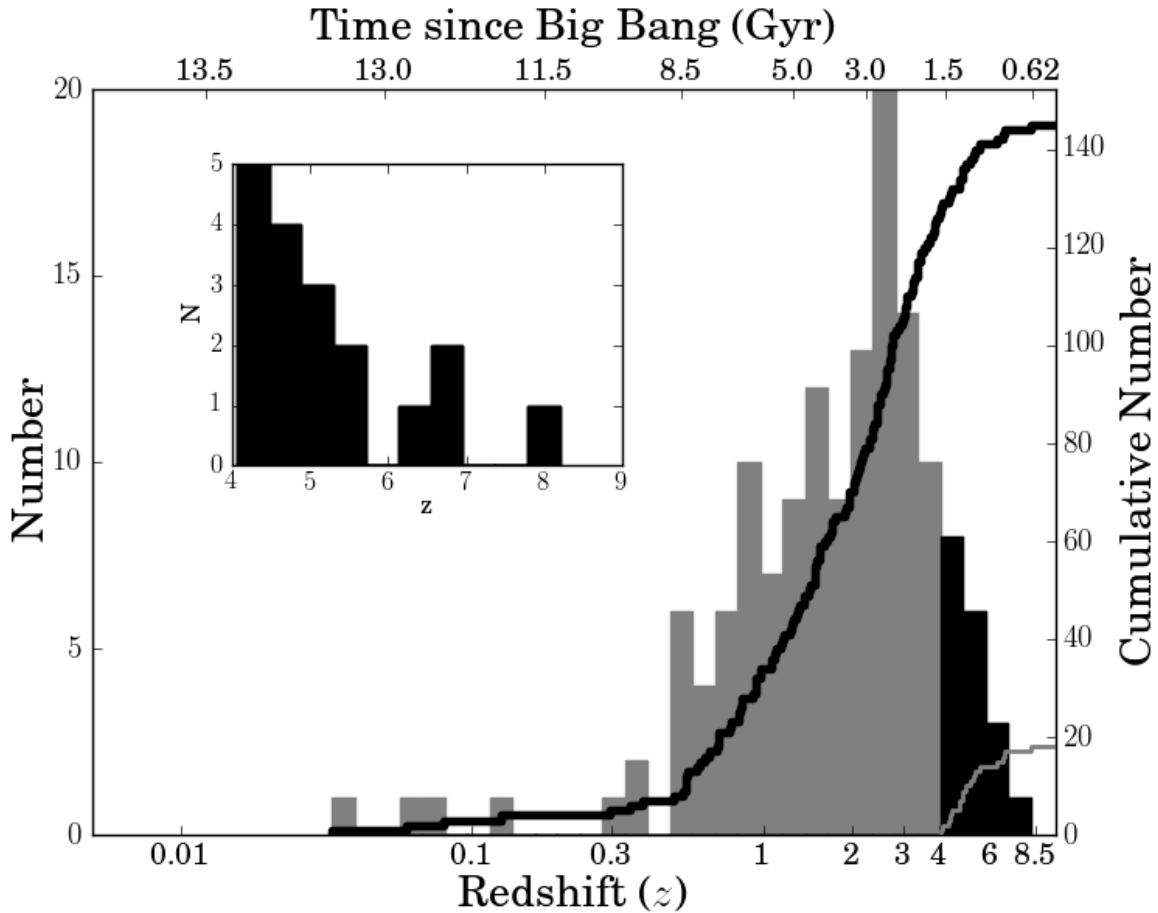}}
\  \caption[Redshift distribution of long-duration \Swift{}  GRBs.]
  {Redshift distribution of the 135 long-duration \Swift{} GRBs in our sample (Table \ref{tab:trainingredshifts}).  For the purposes of 
this study, ``high'' redshift is defined as those bursts with redshifts larger than $z=4$, which corresponds to approximately 1-$\sigma$ above the mean of the distribution.  In our sample, 18 bursts 
fall into this category (black, and in inset). In determining age since the Big Bang, we assume a cosmology with $h=0.71$, $\Omega_m=0.3$, and $\Omega_\Lambda=0.7$. Solid lines show the cumulative number of GRBs as a function of redshift for high-$z$ bursts (grey) and all bursts (black).
 }
\label{fig:reddist}
\end{figure}

As the most luminous electromagnetic explosions, gamma-ray bursts (GRBs) offer a unique probe into the distant universe---but only if their rapidly fading afterglows are observed before dimming beyond detectability \citep[e.g.,][]{wijers98,escude98,lamb00,kawai08,McQuinn08}.  Since the launch of the \Swift{} satellite in November 2004 \citep{gehrels04}, more than 170 long duration \Swift{} gamma-ray bursts have had measured redshifts, but only a handful fall into the  highest redshift range that allow for the probing of the earliest ages of the universe, up to less than a billion years after the Big Bang (Fig. \ref{fig:reddist}).  With a limited budget of large-aperture telescope time accessible for deep follow-up, it is becoming increasingly important to rapidly identify these GRBs of interest in order to capture the most interesting events without spending available resources on more mundane events. 

Along with quasars \citep[e.g.,][]{Mortlock11} and NIR-dropout lyman-break galaxies \citep[e.g.,][]{Bouwens10,Bouwens11}, GRBs have been established as among the most distant objects detectable in the universe, with a spectroscopically confirmed event at $z=8.2$
\citep[GRB 090423;][]{tanvir09a,salvaterra09a} and a photometric candidate at $z \sim 9.4$ \citep[GRB 090429B;][]{Cucchiara11}.  
Such observations can provide valuable constraints on star formation in the early universe, illuminate the locations and properties of some of the earliest galaxies and stars, and probe the epoch of reionization. \citep[e.g.,][and references therein]{Tanvir07}.
Further, the relatively simple spectra of GRB afterglows compared to other cosmic lighthouses makes it easier to both identify their redshifts and extract useful spectral features such as neutral hydrogen absorption signatures for the study of cosmic reionization. \citep[e.g.,][]{escude98,Barkana04,Totani06,McQuinn08}.  However, such benefits can only be realized if spectra are obtained with large-aperture telescopes before the afterglow fades beyond the level required to obtain a useful signal, typically within a day after the GRB.

As such, there has been a long-standing effort to extract a measure of a GRB's redshift from its early time, high-energy signal, with a primary goal of the rapid identification of high-$z$ candidates. 
This might appear in principle to be a straightforward exercise; for instance, distant GRBs should on average appear fainter and longer-duration than nearby events due to distance and cosmological time dilation, respectively.  In practice, however, the large intrinsic diversity of GRBs, as well as thresholding effects, confounds the straightforward use of early-time observations in divulging redshift and other important properties.  While much effort has gone into tightening the correlations between high-energy properties in order to homogenize the sample for use as a luminosity (and hence distance/redshift) predictor \citep[e.g.,][]{amati02,ghirlanda04,firmani06,schaefer07}, there has been significant debate as to whether some of these relations are actually due to thresholding effects specific to the detectors rather than intrinsic physical properties of the GRBs \citep[e.g.,][]{friedman05,butler07b,butler09}.  Regardless, whether or not these inferred relationships are actually physical or simply detector effects would not affect their utility as a {\it detector-specific} parameter prediction tool.  By restricting ourselves to \Swift{} events only, we avoid the uncertainty of whether certain correlations remain when using different detectors.

With this in mind, we set out to search for indications of high-redshift GRBs in the rich, mostly homogeneous dataset provided by 6+ years of GRB observations by the three telescopes onboard \Swift{}  (BAT; \citealt{barthelmy05}, XRT; \citealt{burrows05}, UVOT; \citealt{roming05}).  Past studies exploring high-$z$ indicators have used hard cuts on certain features such as UVOT afterglow detection, burst duration, and inferred hydrogen column density \citep[e.g.,][]{grupe07,vandenberk08,ukwatta09}, regression on such features \citep{koen09,koen10}, and combinations of potential GRB luminosity indicators \citep{xiao09,xiao11}. In this work, we take a different approach by utilizing supervised machine learning algorithms, specifically Random Forest classification, to make follow-up recommendations for each event automatically and in real time. Particular attention is paid to careful treatment of performance evaluation by using cross-validation (\S\ref{sec:results}), a robust methodology to guard against over-fitting and the circular practice of testing hypotheses using the same data that suggested (and constrained) them.

The primary driving force of this study is simple: \emph{given limited follow-up time available on telescopes, we want to maximize the time spent on high-$z$ GRBs}\footnote{For the purposes of this study, ``high-redshift'' corresponds to all $z>4.0$: a compromise between only keeping the most interesting events and having enough data to train on.  However, we have explored performance of different redshift cuts; see \S\ref{sec:finalclassifier}.}. To this end, we provide a deliverable metric, explained in \S\ref{sec:deliverables}, to assist in the decision making process on whether to follow up a new GRB.  Real-time distribution of this metric is available for each new \Swift{}  trigger via website\footnote{\url{http://rate.grbz.info/}} and RSS feed\footnote{\url{http://rate.grbz.info/rss.xml}}.

The structure of this paper is as follows: in \S\ref{sec:obs} we outline the collation of the data, and describe the particular GRB features utilized in redshift classification.  In \S\ref{sec:methods}, the Random Forest algorithm is detailed, along with some specific challenges posed by this particular data set.  Performance metrics of the classifiers are presented in \S\ref{sec:results}, and in \S\ref{sec:application} we discuss the results of testing the classifiers on additional GRBs, both with and without known redshifts.  Finally, our conclusions are given in \S\ref{sec:conclusions}.

\section{Data Collection}
\label{sec:obs}

The \Swift{}  BAT constantly monitors 1.4 steradians on the sky over the energy range $15-150$ keV.  GRB triggering can occur either by a detection of a large gamma-ray rate increase in the BAT detectors (``rate trigger''), or a fainter, long-duration event recovered after on-board source reconstruction reveals a new significant source (``image trigger'').  A rough ($\sim$ 3 arcmin) position is determined, and if there are no overriding observing constraints, the spacecraft slews to allow the XRT and UVOT to begin observations, typically between 1 and 2 minutes after the trigger.  The XRT observes between the energy range of $0.2-10$ keV and detects nearly all of the GRBs it can observe rapidly enough, providing positional accuracies of $2-5$ arcseconds within minutes. The UVOT is a 30cm aperture telescope that can observe in the range of $170-650$ nm. Due to the relatively blue response of this telescope, it cannot detect highly reddened sources due to either dusty environments or (more relevant to this analysis) high-redshift origins.

At each stage in the data collection process, information is sent to astronomers on the ground via the Gamma-ray bursts Coordinates Network (GCN\footnote{\url{http://gcn.gsfc.nasa.gov/}}) providing rapid early-time metrics.  The more detailed full data are sent to the ground in $\sim$ 90 minute intervals starting between roughly $1-2$ hours after the burst.  For our dataset, we have collected data after various levels of processing directly from GCN notices, online tables\footnote{\url{http://swift.gsfc.nasa.gov/docs/swift/archive/grb\_table.html/}} and automated  pipelines \citep{butler07a,butler07b} that process and refine the data into more useful metrics.  Tens of attributes and their estimated uncertainties (when available) are parsed from the various sources and collated into a common format.  

In order to evaluate our full dataset in an unbiased way, we restricted ourselves to using features which have been generated for all possible\footnote{Even with the restriction of observation by all 3 \Swift{}  telescopes, certain features derived from model fits are nonetheless incalculable for certain GRBs from the available data. See \S\ref{sec:missing} for how our algorithm treats missing values.} past events and are automatically generated for future events.  This is the primary reason we do not include potentially useful features such as relative spectral lag \citep[e.g.,][and references therein]{ukwatta10,ukwatta11} which has been utilized as a redshift indicator with smaller and pre-\Swift{} datasets \citep{murakami03,band04,zhang06,schaefer07} but requires a larger spectral coverage than \Swift{} alone can provide. However, our technique is easily extendable to include additional useful features should they be homogeneously determined for past GRBs and automatically available in real-time for new events, and therefore we strongly encourage the automated distribution of any such data products.

Because the addition of too many features causes a decrease in classifier performance (see \S\ref{sec:comparesets}), a total of 12 features were kept for our final classifier (Table \ref{tab:features}), 10 of which were derived from BAT gamma-ray measurements, one from XRT observations, and one from UVOT observations.  Of the 10 BAT features, 4 were parsed directly from GCN Notices, the most rapidly available (and thus unrefined) source of information on GRBs\footnote{For 14 events in our test set, the SWIFT\_BAT\_POSITION notice was not available on the online repository, primarily due to satellite downlink problems at the time of discovery. For these events, the relevant parameters were extracted directly from the \Swift{} TDRSS database (\url{http://heasarc.nasa.gov/W3Browse/all/swifttdrss.html}).}.  The parameter $t_{BAT}$ is a rough measurement of the duration of the BAT trigger event and thus a lower limit on the total duration of the GRB.  The binary feature of whether or not the event was a rate trigger is an indicator of the signal-to-noise of an event, for only the brighter events are detected as rate triggers, while those on the threshold of detection are image triggers.  The final two GCN features are also rough indicators of brightness:  $\sigma_{BAT}$ is the significance (in sigma) of the detected source in the on-board reconstruction of the BAT image, and $R_{peak,BAT}$ is the peak count rate observed during the duration of the event.
 
\begin{deluxetable}{lll}	
\tabletypesize{\scriptsize}
\singlespace
\tablewidth{0pt}
\tablecaption{List of Features Utilized }
\tablehead{\colhead{Feature} &
\colhead{Type} &
\colhead{Reference}  
 }
\startdata
 BAT Rate Trigger? 	& BAT Prompt 	& GCN Notices  \\
 $\sigma_{BAT}$		& BAT Prompt	& GCN Notices	 \\
 $R_{peak,BAT}$		& BAT Prompt	& GCN Notices	 \\
 $t_{BAT}$			& BAT Prompt	& GCN Notices	 \\
 UVOT Detection? 	& NFI Prompt 	& GCN Notices 	 \\
 $N_{\rm H,pc}$		& Processed  	& \citet{butler07a}		 \\
 $\alpha$			& Processed		& \citet{butler07b}		 \\
 $E_{\rm peak}$		& Processed		& \citet{butler07b}		 \\
 $S$				& Processed		& \citet{butler07b}		 \\
 $S/N_{\rm max}$	& Processed		& \citet{butler07b}		 \\
 $T_{90}$			& Processed 	& \citet{butler07b}		 \\
 $P_{z>4}$			& Processed		& \citet{butler10}		

\enddata
\label{tab:features}
\end{deluxetable}

Five higher-level BAT-derived attributes were pulled from online tables automatically updated by the pipeline described in \citet{butler07b}.  The feature $\alpha$ is the power-law index before the peak of the Band-function fit to the  gamma-ray spectrum (typically clustered around $-1$). Another parameter in the Band-function fit, $E_{\rm peak}$, is the energy at which most of the photons are emitted. The fluence, $S$, is the total gamma-ray flux (15--350 keV) integrated over the duration of the burst.  $S/N_{\rm max}$ is simply the maximum signal-to-noise achieved over the duration of the light curve. Finally, $T_{90}$ is a measure of the burst duration, defined to be the time interval over which the middle 90\% of the total background-subtracted flux is emitted.

One additional ``metafeature'' is derived from the BAT data. In principle, if we knew in detail the intrinsic distributions of GRB
observables \citep[fluence, hardness, duration; see][]{butler07b} as a function of redshift,
measurements of these observables for a new event could be used to
directly evaluate the expected redshift.  A detailed fitting
of the intrinsic distributions for Swift is presented in \citet{butler10}, and we use the parametrized intrinsic distributions
there to calculate the posterior probability redshift distributions for
each GRB in our sample \citep[see, e.g., Figure 8 in][]{butler10}.
Here, we further condense this distribution into one useful feature: $P_{z>4}$, the fraction of posterior probability at $z>4$.

Finally, two features are extracted from data taken by the two narrow-field instruments onboard \Swift, one each from the XRT and UVOT.  The feature $N_{\rm H,pc}$ is the excess neutral hydrogen column (above the galactic value) inferred from the XRT PC (Photon-counting mode) data, obtained from the \citet{butler07a} pipeline.  The last feature is simply a binary measure of whether or not the GRB afterglow was detected by the UVOT.  

While most of these features have associated uncertainties, the proper treatment of uncertainties in attributes is an area of ongoing research in machine learning \citep[e.g.][]{carroll06}. Some methods call for the uncertainties to be treated as attributes in and of themselves, but we found that the addition of these relatively weak features were actually detrimental for our small dataset (see, e.g., Fig. \ref{fig:useless}). We also considered an approach by which features with large uncertainties were considered poor measurements and were instead marked as missing values.  However, this had a negligible effect on our final classifier performance, so for simplicity we treat all values as precisely known.  

\begin{figure}[t!]
  \centerline{\plotone{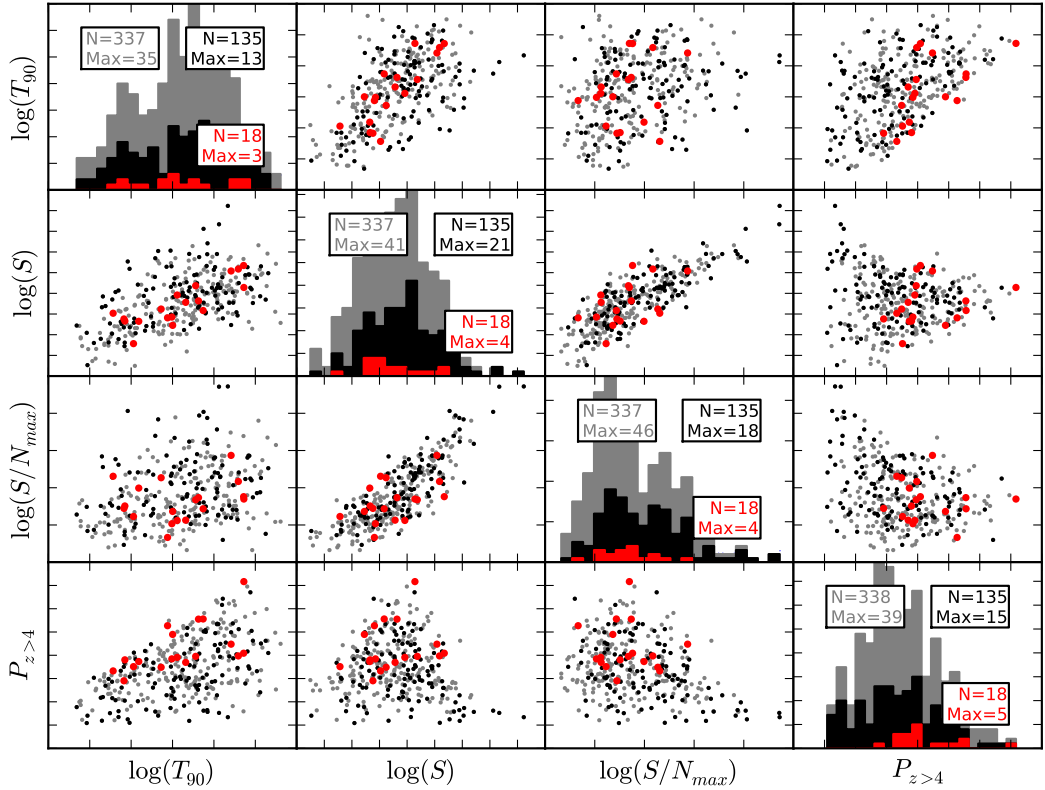}}
\  \caption[?????]
  {Plot of a selection of early-time \Swift{} features (Table \ref{tab:features}) against each other. The grey points show the full distribution of Swift GRBs. Bursts with known redshifts are black, and the 18 known events with redshifts greater than 4 are overplotted in red. In the histogram text boxes, $N$ shows how many instances of that feature in total are shown (anything less than the full number of instances is due to the value of that feature being unknown for certain instances), and Max shows the maximum number of instances in any particular bin. 
 }
\label{fig:featuresvfeatures}
\end{figure}

We collated data on all \Swift{} GRBs with rapidly available BAT data up to and including GRB 100621A - 471 in total.   Specifically, this excludes bursts which were not identified in real-time due to the event being below the standard triggering threshold or occurring while the satellite was slewing to a new location.
Of these, 39 are short GRBs (defined for the purposes of this study to 
be those with $T_{90} < 2.0$ s\footnote{
$T_{90}$ alone is not a strong enough discriminator to definitively assign a particular GRB to one class or another (``short'' versus ``long''; see \citealt{levesque10} for discussion). In this study, we will accept the few errant bursts from the ``short'' class included in our sample as additional noise in our method.}), which are believed to arise from a different 
physical process and are thus removed from the sample. 
For further uniformity in the sample, bursts without rapid ($< 1$ hour)
XRT/UVOT follow-up are also removed, leaving 347 events\footnote{The reason for this missing data is almost always due to observing constraints from the GRB being too close to the Sun, Moon, or Earth at the time of discovery.  Not removing these bursts would introduce a bias in the sample due to the fact that events without a rapid XRT position are far less likely to lead to an afterglow discovery, and hence, redshift determination. A total of 15 bursts with known-$z$ were removed because of this.}. Of the remaining long bursts in our sample, 135 had reliable redshifts (Table \ref{tab:trainingredshifts}) and were thus included in our training data set (Table \ref{tab:training}).  The additional 212 long bursts without secure redshift determinations are explored further in \S \ref{sec:unknownz}.  Exploratory data analysis shows preliminary indications of which of these features will be most useful for classification.  Figure \ref{fig:featuresvfeatures} shows several 2D slices of the feature space, with the high-$z$ bursts highlighted. 

\begin{deluxetable}{llll}
        \tabletypesize{\scriptsize}
        \singlespace
        \tablewidth{0pt}
        \tablecaption{Training Data Redshifts}
        \tablehead{\colhead{GRB} &
\colhead{$\widehat{\mathcal{Q}}_{train}$} &
\colhead{$z$} &
\colhead{References} }
\startdata
050223	&	4.30e-01	&	0.5915	&	\citealt{berger06a} \\ 
050315	&	3.57e-01	&	1.949	&	\citealt{kelson05a} \\ 
050318	&	6.86e-01	&	1.44	&	\citealt{berger05b} \\ 
050319	&	5.90e-01	&	3.2425	&	\citealt{fynbo05a,jakobsson06a,fynbo09} \\ 
050416A	&	7.68e-01	&	0.6535	&	\citealt{cenko05a} 
\enddata
\tablecomments{Table \ref{tab:trainingredshifts} is published in its entirety in the electronic edition of The Astrophysical Journal. A portion is shown here for guidance regarding its form and content.}
\label{tab:trainingredshifts}
\end{deluxetable}

\begin{landscape}
            \begin{deluxetable}{lllllllllllll}
        \tabletypesize{\scriptsize}
        \singlespace
        \tablewidth{0pt}
        \tablecaption{Training Data}
        \tablehead{\colhead{GRB} &
\colhead{$\alpha$} &
\colhead{$E_{peak}$} &
\colhead{$S$} &
\colhead{$S/N_{max}$} &
\colhead{$N_{H,pc}$} &
\colhead{$T_{90}$} &
\colhead{$\sigma_{BAT}$} &
\colhead{$R_{peak,BAT}$} &
\colhead{Rate} &
\colhead{$t_{BAT}$} &
\colhead{UVOT} &
\colhead{$P_{z>4}$} \\
\colhead{} &
\colhead{} &
\colhead{(keV)} &
\colhead{(erg/cm$^2$)} &
\colhead{} &
\colhead{$(10^{22}$ cm$^{-2})$} &
\colhead{(s)} &
\colhead{} &
\colhead{(ct/s)} &
\colhead{trigger} &
\colhead{(s)} &
\colhead{detect} &
\colhead{} }
\startdata
050223	&	-1.74e+00	&	6.70e+01	&	8.75e-07	&	1.34e+01	&	-2.37e-01	&	1.74e+01	&	9.00e+00	&	7.26e+02	&	yes	&	8.19e+00	&	no	&	1.74e-01 \\ 
050315	&	?	&	4.33e+01	&	4.32e-06	&	4.37e+01	&	9.60e-02	&	9.46e+01	&	8.00e+00	&	2.60e+02	&	yes	&	1.02e+00	&	no	&	9.27e-02 \\ 
050318	&	-1.22e+00	&	5.01e+01	&	1.41e-06	&	4.90e+01	&	1.80e-02	&	3.10e+01	&	9.00e+00	&	2.05e+02	&	yes	&	5.12e-01	&	yes	&	6.29e-02 \\ 
050319	&	-2.00e+00	&	4.47e+01	&	1.87e-06	&	1.82e+01	&	1.50e-02	&	1.54e+02	&	1.00e+01	&	2.63e+02	&	yes	&	1.02e+00	&	yes	&	1.48e-01 \\ 
050416A	&	-7.24e-01	&	1.50e+01	&	3.40e-07	&	1.75e+01	&	2.34e-01	&	2.91e+00	&	1.10e+01	&	1.65e+02	&	yes	&	5.12e-01	&	yes	&	4.35e-03 
\enddata
\tablecomments{Table \ref{tab:training} is published in its entirety in the electronic edition of The Astrophysical Journal. A portion is shown here for guidance regarding its form and content.}
\label{tab:training}
\end{deluxetable}
\end{landscape}

\section{Classification Methodology}
\label{sec:methods}

The resource allocation approach we have taken here naturally manifests itself as a classification problem: deciding whether or not to follow up a new event is simply a two-class problem of ``observe'' or ``do not observe,'' and the methodology presented here can be applied to any problem that can be broken up in this way. This was the primary motivation of using classification instead of a regression or ``pseudo-$z$'' approach for this study.  
The primary disadvantage of classification for the particular problem of high-redshift identification is that all instances above and below the class division
(chosen here to be $z=4$) are treated equally; e.g., a burst with $z=4.01$
has the same influence on our inference about ``high'' bursts as a burst
with $z=8$\footnote{This of course would not be an issue when applying the \rate{} methodology to a problem with more well-defined class boundaries, such as prioritizing follow-up of a particular rare class of transient event.}.
However, classification has advantages over regression in that it is a conceptually much simpler problem, and most of the difficulties encountered due to the unbalanced, small dataset of interest here would only be aggravated by an extension to regression.  Further, our approach capitalizes on the fact that one of our predictors (lack of UVOT detection) is itself a binary feature with an understood physical connection to redshift\footnote{Bursts with a UVOT detection must be $z <~ 5$ due to the Lyman cutoff. This is due to the fact that photons with wavelengths smaller (thus higher energy) than the Lyman limit of $\lambda=912$\AA{} would be almost completely absorbed by neutral gas in the host galaxy and intergalactic star forming regions. A redshift of $z=5$ might therefore be considered a natural cutoff point for the high-$z$ class, but due to so few training events at this high redshift ($N_{z>5} = 8$), we opted for the more conservative cutoff point of $z=4$ ($N_{z>4} = 18$).}.

\subsection{Random Forest classification} \label{sec:rf_class}

A supervised classification algorithm uses a set of training data of known class to estimate a function for assigning data points to classes based on their features. The statistics and machine learning communities have developed many classification algorithms, including Support Vector Machines (SVM), Na\"ive Bayes, Neural Networks, and Gaussian Mixture Models. We use Random Forest \citep[RF][]{breiman:2001:random} for its ability to select important features, resist overfitting the data, model nonlinear relationships, handle categorical variables, and produce probabilistic output.  
These strengths, along with a record of attaining very high classification accuracy relative to other algorithms have led to widespread use of Random Forest in the astronomy community~ \citep[e.g.,][]{bailey:2007:find,carliles:2010:random,dubath11,okeefe:2009:star,richards11}. In this work, we utilized custom \texttt{R} software built around the \texttt{randomForest} package to generate classifiers and evaluate performance. 

Random Forest is an ensemble classifier that averages together the outputs from many decision trees, a common example of which is Classification and Regression Trees \citep[CART,][]{breiman84}. In RF, the decision trees are constructed by recursive binary splitting of the high-dimensional feature space, where each split is performed with respect to a particular feature. For example, the decision tree might split the data on feature $S/N_{\rm max}$ using value $100$, in which case all observations with $S/N_{\rm max} > 100$  are placed in one group and the rest placed in the second group. As these are binary splits, for convenience we henceforth refer to observations going ``left'' or ``right'' of each split as an analogue for the decision made at that split.

For each split, the feature and specific split-point are chosen so as to best separate the observations into the classes, by using some objective function. We use the Gini Index, a standard objective function for classification \citep{breiman84}. At any given node in a tree and some proposed split $s$, let $N_{l,h} =$ number of high-priority (in our case, high-$z$) events that go to the left of the split, $N_{l,l} = $ number of low-priority events that go left. Define $N_{r,h}$ and $N_{r,l}$ similarly, replacing left with right. Let $N_{l} = N_{l,l} + N_{l,h}$, the total number of observations that go left. Similarly define, $N_{r} = N_{r,l} + N_{r,h}$, for the total number of observations that go right. The Gini criterion is defined as
\begin{equation}
\label{eqn:gini}
\frac{N_l}{N_l + N_r} \left( \frac{N_{l,h}}{N_l} \right) \left( \frac{N_{l,r}}{N_l} \right) + \frac{N_r}{N_l + N_r} \left(\frac{N_{r,h}}{N_r}\right)\left(\frac{N_{r,l}}{N_r}\right),
\end{equation}
and the split that minimizes this value over the random subset of features considered at each node\footnote{At each node, $m=3$ features were considered, guided by the default practice in the \texttt{randomForest} routine of $m = \mbox{floor}(\sqrt{p})$, where $p$ is the total number of features.} is chosen.  For instance, in the ideal case where the split on a particular feature completely separates all the instances of the two classes from each other, the Gini index reaches a minimum of 0.
The splitting is done recursively, continuing down each subgroup until
all of the observations in each final group (``terminal node'') are of a single class.
The process is known as ``growing a tree'' because each split can be visualized as generating two branches from a single branch to produce a tree-like structure. Once a tree is constructed from the training data, each new
observation starts at the root node (the top split in the tree) and,
recursively, the splitting rules determine the terminal node to which
the observation
belongs. The observation is assigned to the class of the terminal node.

To create the RF classifier, a sufficiently large\footnote{With enough trees, error rates will converge and growing additional trees will result in no further performance improvements. Our forests are grown to 5000 trees throughout this work in order to ensure consistency in the rankings of unknown events.} number of decision trees are constructed, resulting in a ``forest''.
Each decision tree is generated from an independent bootstrap sample \citep{bootstrap}; Samples are drawn with replacement from the original data set, resulting in a new data set of the same size as the original, with on average 2/3 of the original observations present at least once.
Additionally, only a random subset of the features is eligible for splitting at each node. Many decision trees are grown with each tree slightly different due to the bootstrap sampling and random selection of features at each split. RF classifies new observations by averaging the outputs of each tree in the ensemble. 

Training observations can be classified by using all trees where that observation was not used in the bootstrap sampling stage. This produces estimates of error rates and class probabilities for each observation that are not overfit to the training data. Error rates and probabilities computed using this method are known as ``out-of-bag'' estimates.

\subsubsection{Missing feature values}
\label{sec:missing}

As mentioned in \S\ref{sec:obs}, certain features, namely $\alpha$ and $N_{\rm H,pc}$, were occasionally unable to be determined from model fits to the data and are thus missing for certain observations. 
We handle missing values by imputation, where missing values for features are assigned estimated values.  For missing values of continuous features, we assigned the median of all observations for which that feature is non-missing. Missing categorical features are assigned the mode of all observations for which the feature is non-missing. This is one of the simplest imputation methods and has the advantage of being transparent and computationally cheap. We experimented with a more sophisticated imputation method, \texttt{MissForest}, that iteratively predicts the missing values of each feature given all the other features \citep{missForest}, but as it produced similar error rates to median imputation, we opted for latter, simpler approach in our final classifier.

\subsubsection{Class imbalance} 
\label{sec:imbalance}

A further challenge in this data set is the imbalance between classes. We are training on 135 bursts, only 18 of which are in the high-$z$ class --- an asymmetry present in many resource allocation problems where the goal is to prioritize the rarer events. Without modification, standard machine learning classification algorithms applied to imbalanced data sets attain notoriously suboptimal performance \citep{Chawla:2004:ESI:1007730.1007733}, and often result in simply classifying all unknown events as the more common class.  
As we care more about correctly classifying the rarer events, misclassifications of high-$z$ events must be punished more strongly than vice versa. In Random Forest, classes may be weighted in order to overcome the imbalance by altering the splits chosen by Gini and the probabilities assigned to classes in the terminal nodes of each tree \citep{breimanImbalanceRF}. 

We utilized the \texttt{classwt} option in the \texttt{randomForest} package, which accounts for class weights in the Gini index calculation (Eq. \ref{eqn:gini}) when splitting at the nodes (Liaw 2011, private communication), similar to weighting techniques used in single CART trees \citep{breiman84}.  If we are weighting high-priority observations (e.g. $z > 4$ GRBs) by $w_h$ and low-priority observations by $w_l$, we let,
\begin{align*}
&N^{'}_{l,h} = w_hN_{l,h}\\
&N^{'}_{l,l} = w_lN_{l,l}\\
&N^{'}_{r,h} = w_hN_{r,h}\\
&N^{'}_{r,l} = w_lN_{r,l}
\end{align*}
Let $N^{'}_{l} = N^{'}_{l,l} + N^{'}_{l,h}$, the weighted total number of observations that go left. Similarly define, $N^{'}_{r} = N^{'}_{r,l} + N^{'}_{r,h}$, for the weighted total number of observations that go right. 
The Gini criterion (Eq. \ref{eqn:gini}) is evaluated with the weighted values, and the split that minimizes this value is chosen.
We tested a variety of weight choices by fixing $w_l$ to be unity and varying $w_h$ over a range of values.  The results of this test are presented in \S\ref{sec:weights}, which demonstrates the effects of class weight choice on classifier performance.

\subsection{\rategrbz{} : Random forest Automated Triage Estimator for GRB redshifts} \label{sec:deliverables}

With the background above in hand, we now describe our resource allocation algorithm and its utility for the prioritization of high-$z$ GRB follow-up.   In our application, the data are described in \S\ref{sec:obs} and the classes are high- and low-redshift GRBs, with $z=4$ as the boundary between the classes.
Our primary goal is to provide a decision for each new GRB: should we devote further resources to this event or not? This decision may be different for each astronomer, as it is dependent on the amount of follow-up time available. Implicit in this goal is the desire to follow up on as many truly high-redshift bursts as possible, under a set of given telescope time constraints. 
Directly using the results of an off-the-shelf classifier for this task (i.e., strictly following-up on events labeled as ``high-priority'') is suboptimal. If too few events are labeled as high-priority, there would be an under-utilization of available resources.  If too many are being labeled as high priority, simply following up on the first ones available would preclude any prioritization of events within this high-priority class.  

\begin{figure}[t!]
  \centerline{\plotone{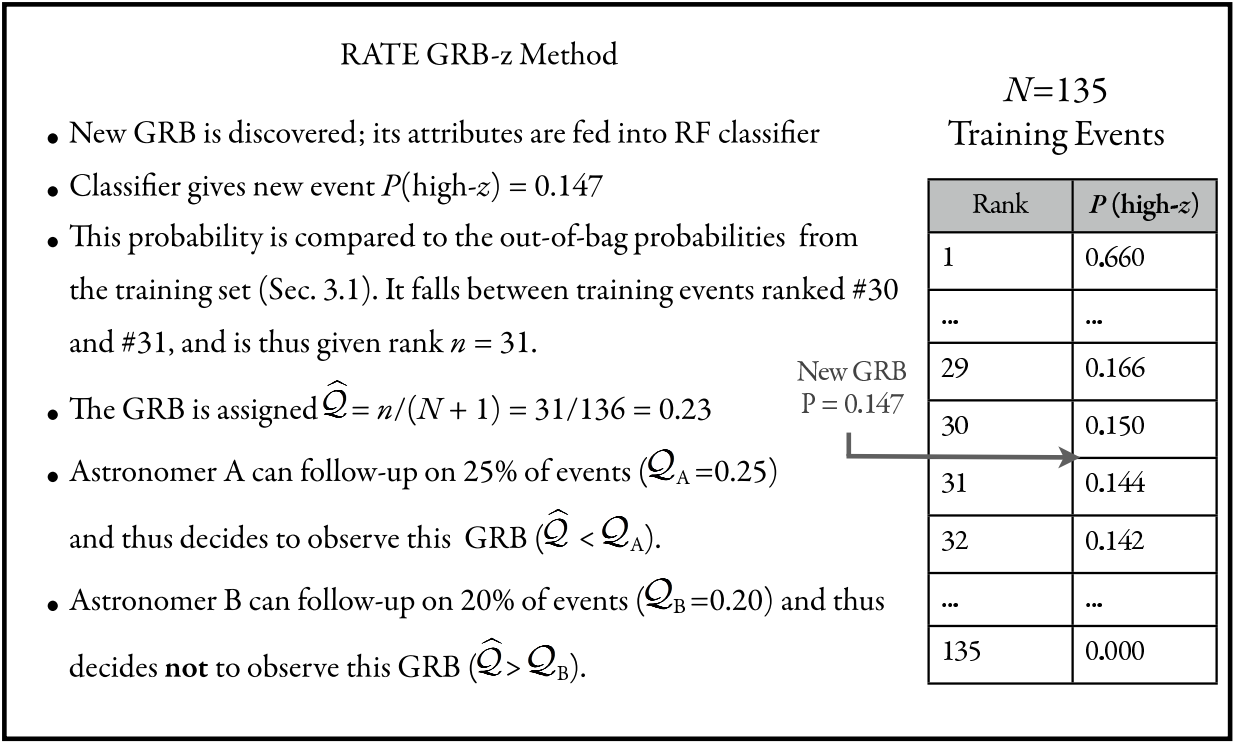}}
\  \caption[Illustration of the RATE Process.]
  {Example of the \rategrbz{} process. 
 }
\label{fig:ratesample}
\end{figure}

These issues can be avoided by instead tailoring the follow-up decision to the resources available (in this case, the available telescope time devoted to high-$z$ GRB observations). The \rate{}  method works as follows: Let $\mathcal{Q}$ be the fraction of events one has resources to follow up on\footnote{As telescope resources are allocated by number of hours and not number of objects, we implicitly assume here that an equal amount of resource time will be allocated to each follow-up event.  This is not in general the case, as objects that turn out to be particularly interesting may have additional resources spent on them. However, a user's estimate of $\mathcal{Q}$ can always be adjusted without penalty as available resources change.}. First we construct a Random Forest classifier using the training data with known response (in this case redshift). We compute the probability of each training event being high-priority using out-of-bag probabilities (See \S\ref{sec:rf_class}). For each new event, we obtain a probability of it being high priority using the Random Forest classifier, and compute the fraction of training bursts that received a higher probability of being high-priority than this new burst. A new burst is assigned rank $n$, with $n-1$ training events having a lower probability of being high priority. Then, for $N$ total training bursts, we obtain a learned probability rank for the new event of  $\widehat{\mathcal{Q}} := n/(N+1)$.  
This leads to a simple decision metric for each new event: If $\widehat{\mathcal{Q}}$ is less than the desired fraction of events a particular observer wishes to follow up ($\widehat{\mathcal{Q}} < \mathcal{Q}$), follow-up observations are recommended.
For instance, if one can afford to follow up on $\sim30\%$ of all observable GRBs, then the desired follow-up fraction is $\mathcal{Q} = 0.3$, and follow-up would be recommended for all events assigned a $\widehat{\mathcal{Q}} < 0.3$.
An illustration of this process in action is shown in Figure \ref{fig:ratesample}. The desired fraction of follow-up events $\mathcal{Q}$ can be dynamically changed without penalty; if the amount of available resources changes, one simply needs to raise or lower this cut-off value accordingly.

\section{Validation of Classifier Performance}
\label{sec:results}

Our training data consist of 135 bursts, 18 of which are high-redshift ($> 4$). Our primary measure of performance is efficiency, defined here as the fraction of high bursts that we that we follow up on relative to the number of total high-z GRBs that occurred ($N_{\rm high~observed}/N_{\rm total~high}$). A secondary performance measure is purity, the number of followed-up events that were actually high-$z$ ($N_{\rm high~observed}/N_{\rm total~observed}$). We measure performance using 10-fold cross-validation ~\citep{kohavi95}, where 90\% of the data is used to  construct a classifier and predict on the remaining 10\% of events.
Each line in the following performance plots is the cross-validated performance averaged across 100 trials of 10-fold cross-validation in order to reduce variability due to randomness in training/test subset selection.

\subsection{Comparison of Weight Choices}
\label{sec:weights}

As described in \S\ref{sec:imbalance}, one of the primary challenges in learning on this dataset is the simple fact that there are comparatively few high-$z$ events on which to train.  If simply getting the most classifications correct were the primary performance metric, as it is in many classification problems, classifying \textit{all} new events as low-redshift would be considered a strong classifier since so few events are in the high-$z$ class.  However, since our objective is to identify the best candidates of this rare class, we punish misclassifications of high-$z$ GRBs more heavily to achieve higher efficiency and purity (outlined above) for a given fraction of followed-up events. 

\begin{figure}[t!]
  \centerline{\plotone{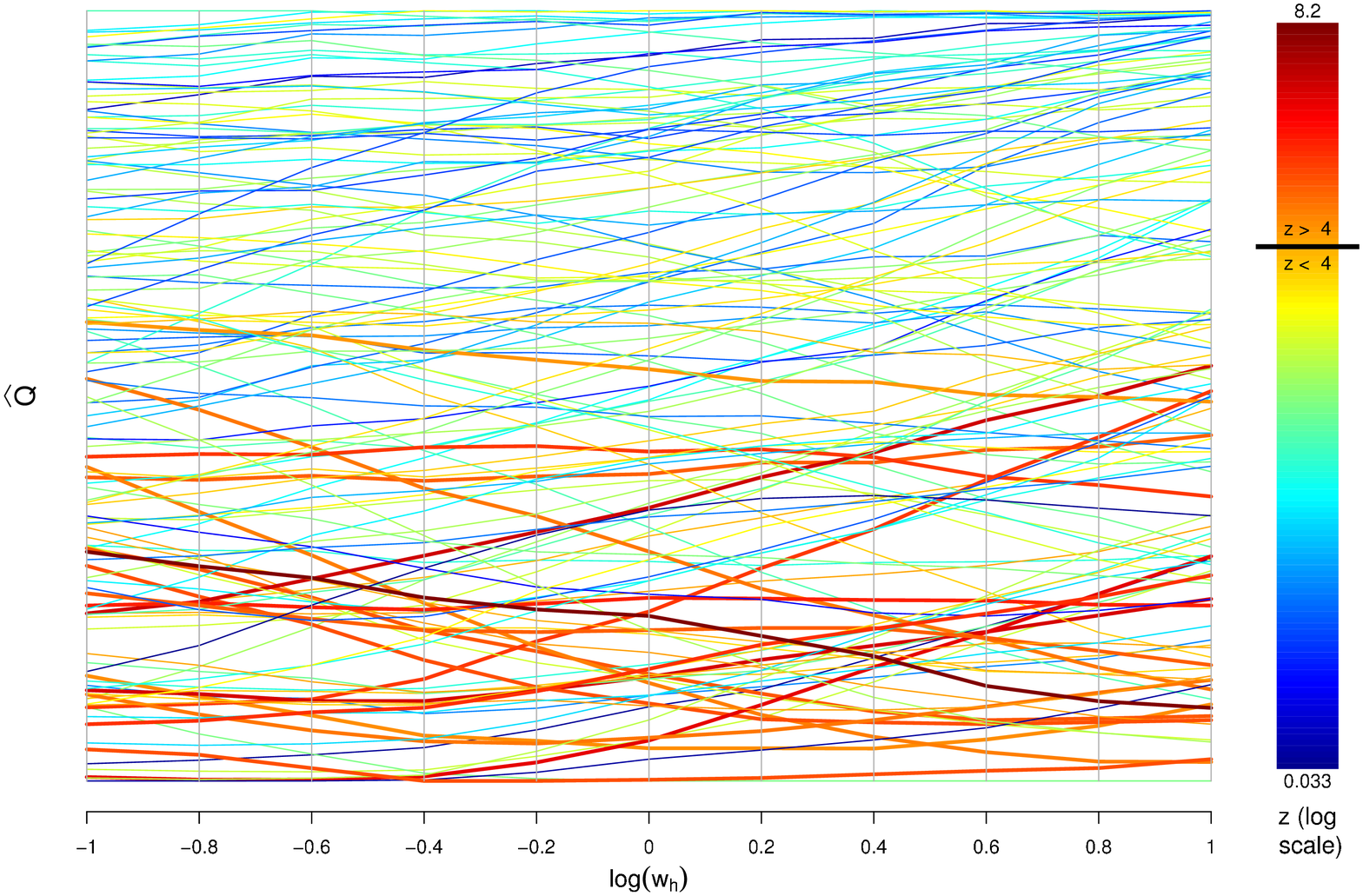}}
\  \caption[?????]
  {Bumps plot showing the cross-validated ranking prediction $\widehat{\mathcal{Q}}$ for each GRB in the training set over a variety of weight choices.  Each line corresponds to an individual GRB, colored by its observed redshift.  Bursts with $z>4$ are plotted with a thicker line.  The clustering of high-$z$ events towards low $\widehat{\mathcal{Q}}$ is clear, illustrating the predictive power of the classifier. The relative ranking of events remains largely stable over different penalization weights, but performance improvements at higher weights are apparent in Figure \ref{fig:weightsefficiency}, which level off after a weight of ~10.}
\label{fig:bumps}
\end{figure}

\begin{figure}[t!]
  \centerline{\plottwo{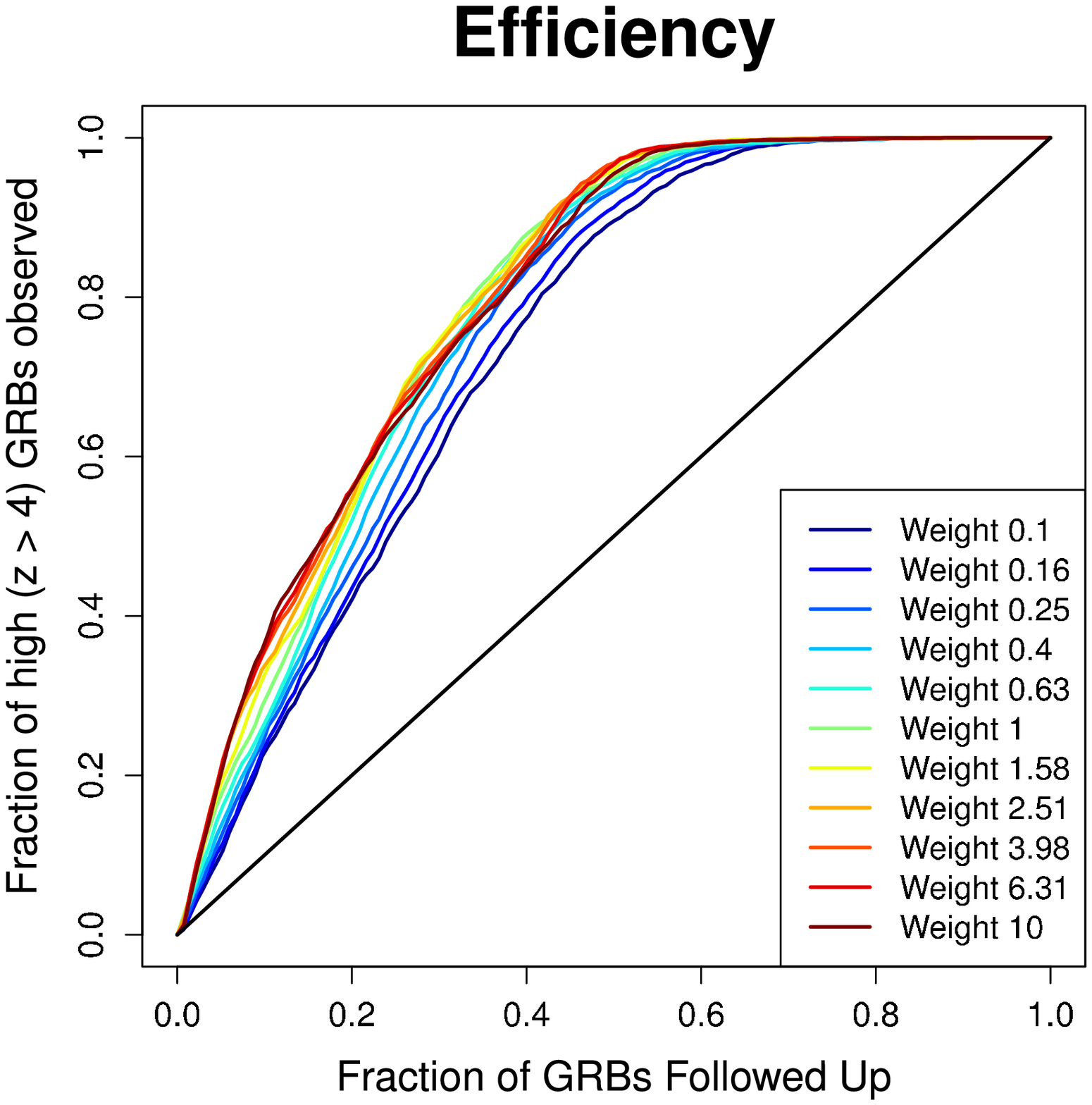}
			{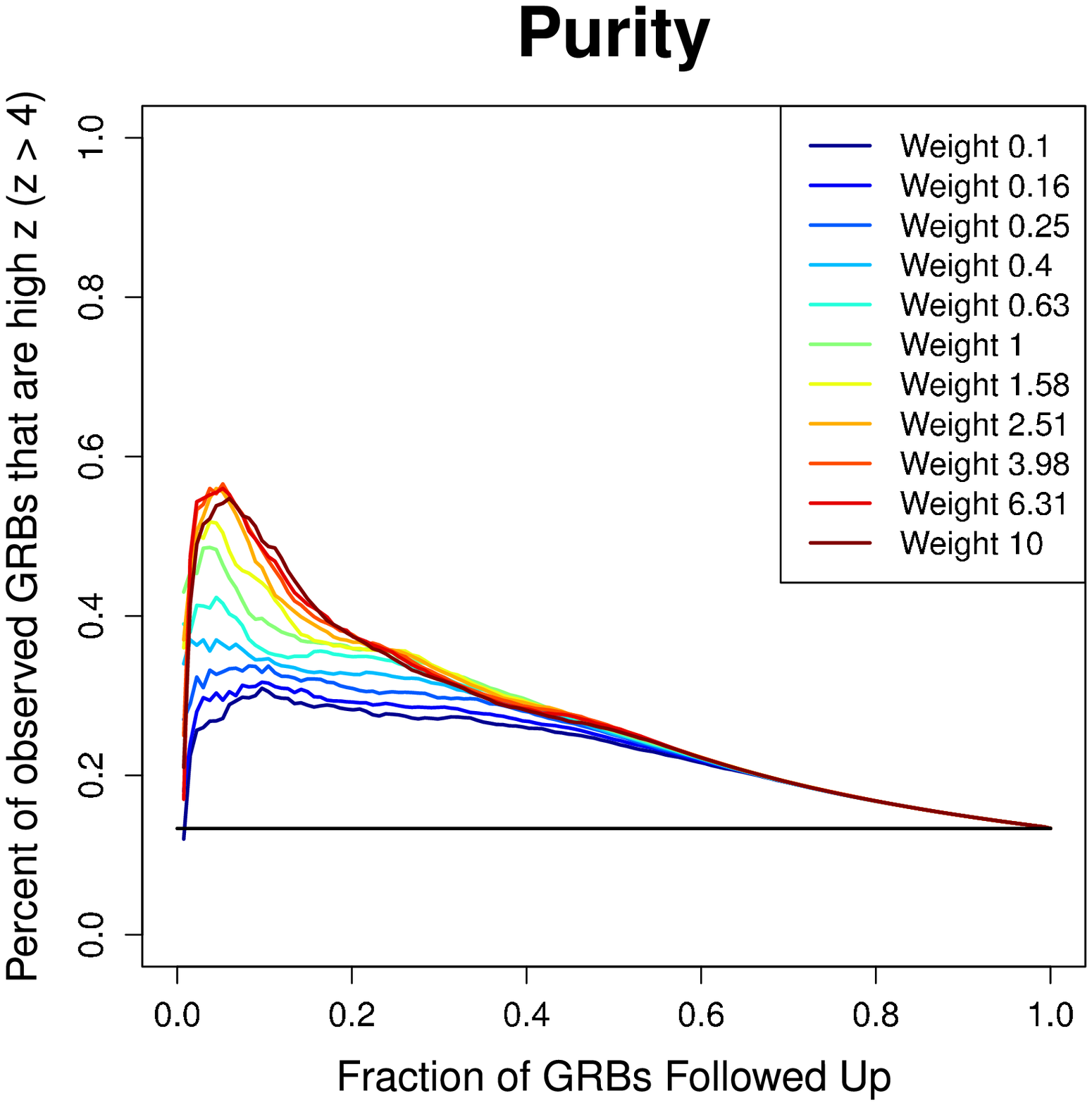}}
\  \caption[?????]
  {The effects of different weights on classifier performance are shown via plots of efficiency ($N_{\rm high~observed}/N_{\rm total~high}$; left panel) and purity ($N_{\rm high~observed}/N_{\rm total~observed}$; right panel) versus fraction of GRBs followed-up (see Figure \ref{fig:popcompare_unknownz} for how our decision criterion $\mathcal{Q}$ corresponds to actual fraction followed-up).  Solid black lines show expected results if selecting events by random guessing alone.  Cross validated performances of the classifier trained with different weights are shown.  Weights above 1.0 penalize misclassifications of high-$z$ events more strongly, and vice versa.  Efficiency and purity were calculated at each fraction of followed-up GRBs ($\mathcal{Q}$, broken down into $N=135$ bins) and averaged over 100 random number generator seeds to account for variance between Random Forest runs.  Clear performance increases for both metrics are shown for higher weights, but beyond a weight of ~10, identical results are achieved.  For clarity, estimates of uncertainties in the curves are not shown, but are of order those plotted in Figure \ref{fig:efficiencypurity4}.
 }
\label{fig:weightsefficiency}
\end{figure}

Thus, in selecting the best weight for our classifier, we compared the efficiency and purity of high-z classification for various  choices of the weight $w_h$ using the feature set shown in Table \ref{tab:features}.  While the relative probability ranking of the GRBs stayed relatively stable over weight choices (Figure \ref{fig:bumps}), a clear trend emerges when comparing classification performance (Figure  \ref{fig:weightsefficiency}). As expected, punishing misclassifications of the smaller, more desirable high-$z$ class cause more of these rare events to be correctly identified.  Beyond a weight of ~10, however, a ceiling is reached where further weight increases show zero change in classification performance. This is therefore the weight chosen for all subsequent performance comparisons.

\subsection{Effects of Feature Selection}
\label{sec:comparesets}
As mentioned in \S\ref{sec:obs}, early testing indicated that the addition of too many features rapidly degraded the predictive power of the final classifier.  This is due to a manifestation of the so-called ``curse of dimensionality'' known as Hughes Phenomenon \citep{hughes68}, where for a fixed number of training instances, the predictive power decreases as the dimensionality increases.  This appears to contradict the conventional wisdom that Random Forest does not overfit, and thus it is better to use many features. However, we note that resistance to overfitting is different from signal being drowned in noise.  With enough noisy features, correlations between class and a useless feature will happen purely by chance, preventing true relationships from being found.  

\begin{figure}[t!]
  \centerline{\plottwo{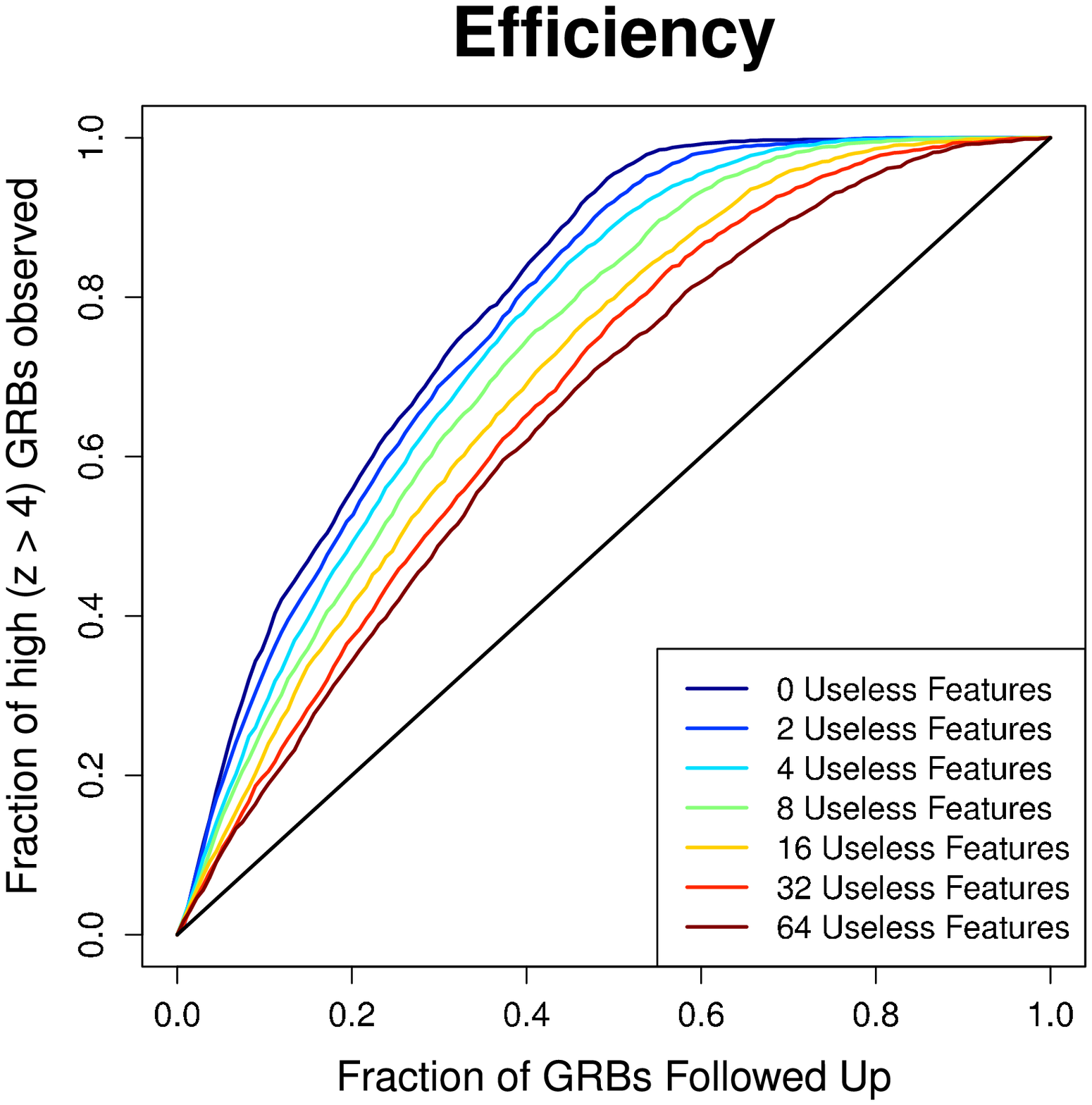}
				{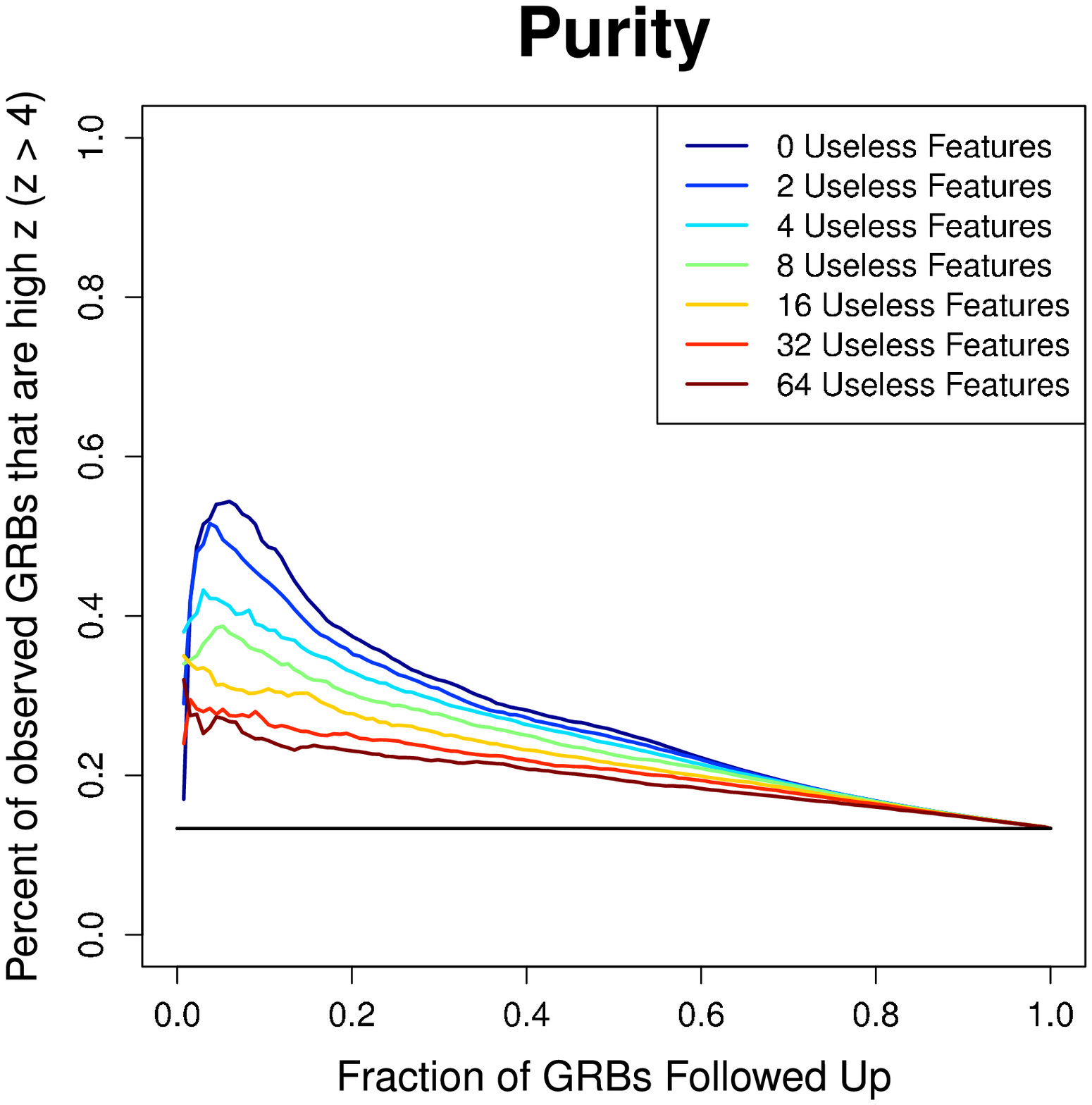}}
\  \caption[?????]
  {The effects of the addition of useless features on classifier performance are shown via plots of efficiency ($N_{\rm high~observed}/N_{\rm total~high}$; left panel) and purity ($N_{\rm high~observed}/N_{\rm total~observed}$; right panel) versus fraction of GRBs followed up according to our decision criterion ($\mathcal{Q}$).  Solid black lines show expected results if selecting events by random guessing alone.  Cross validated performances of the classifier trained with different amounts of useless, randomly generated features are shown.  Degradation in both efficiency and purity becomes clear with the addition of only a few useless features, highlighting the importance of feature selection for small, imbalanced datasets such as this one.}
\label{fig:useless}
\end{figure}

To visualize this effect for our data, we took our nominal feature set and continually added features with no predictive power (random samples from the uniform distribution) to quantify the degradation in performance of the resultant classifiers.  The random features were re-generated for each of the 100 trials, and the cross-validated results are shown in Figure \ref{fig:useless}. The fact that even a small number of useless features causes a noticeable decrease in performance highlights the importance of attribute selection.  However, we note that too much fine tuning of attribute feature selection choices --- such as testing all combinations of features and seeing which one gives the best performance --- would overfit to the data and give an underestimate of the true error.

\subsection{Final Classifier}
\label{sec:finalclassifier}
\begin{figure}[t!]
  \centerline{\plottwo{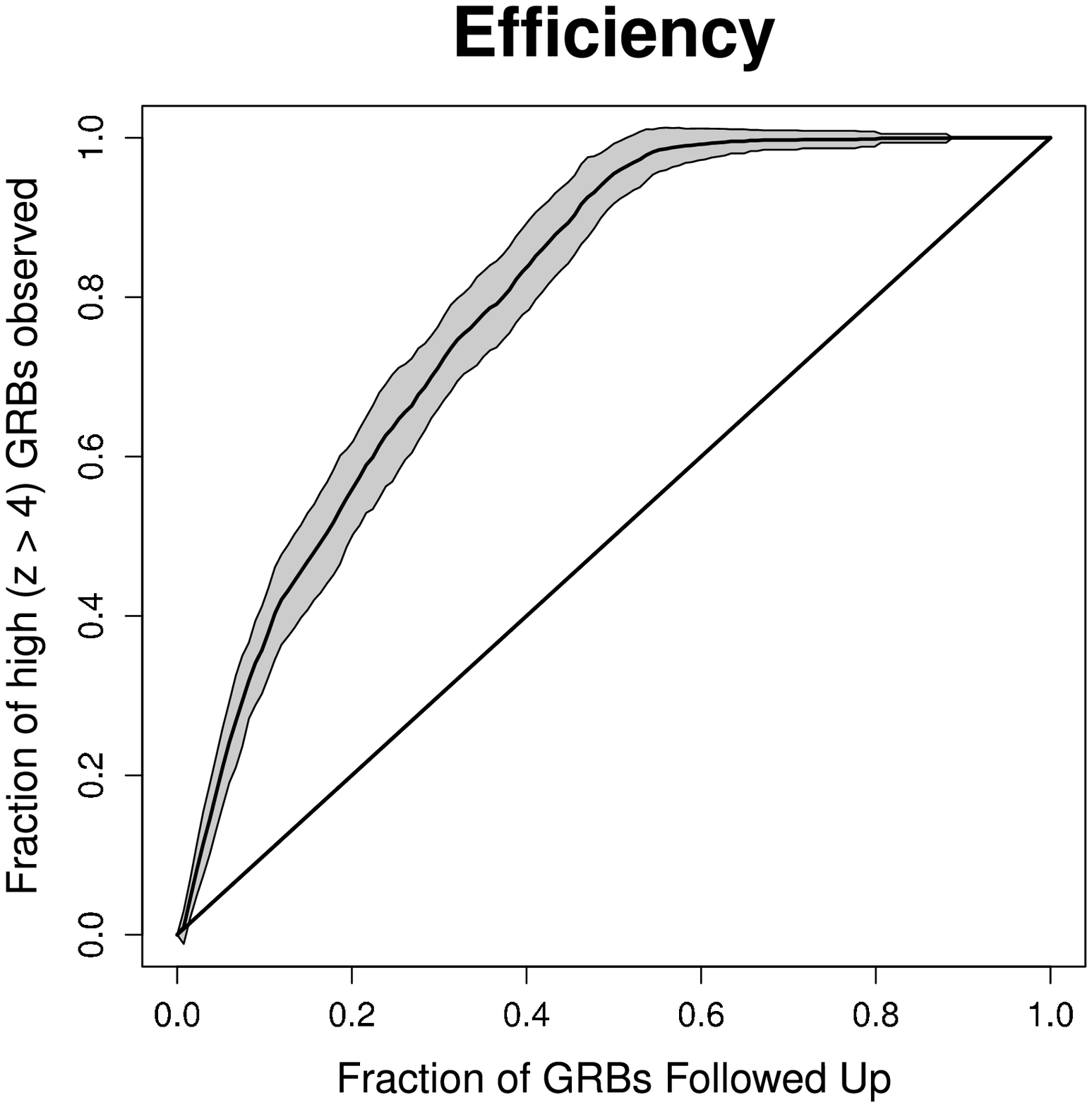}
					{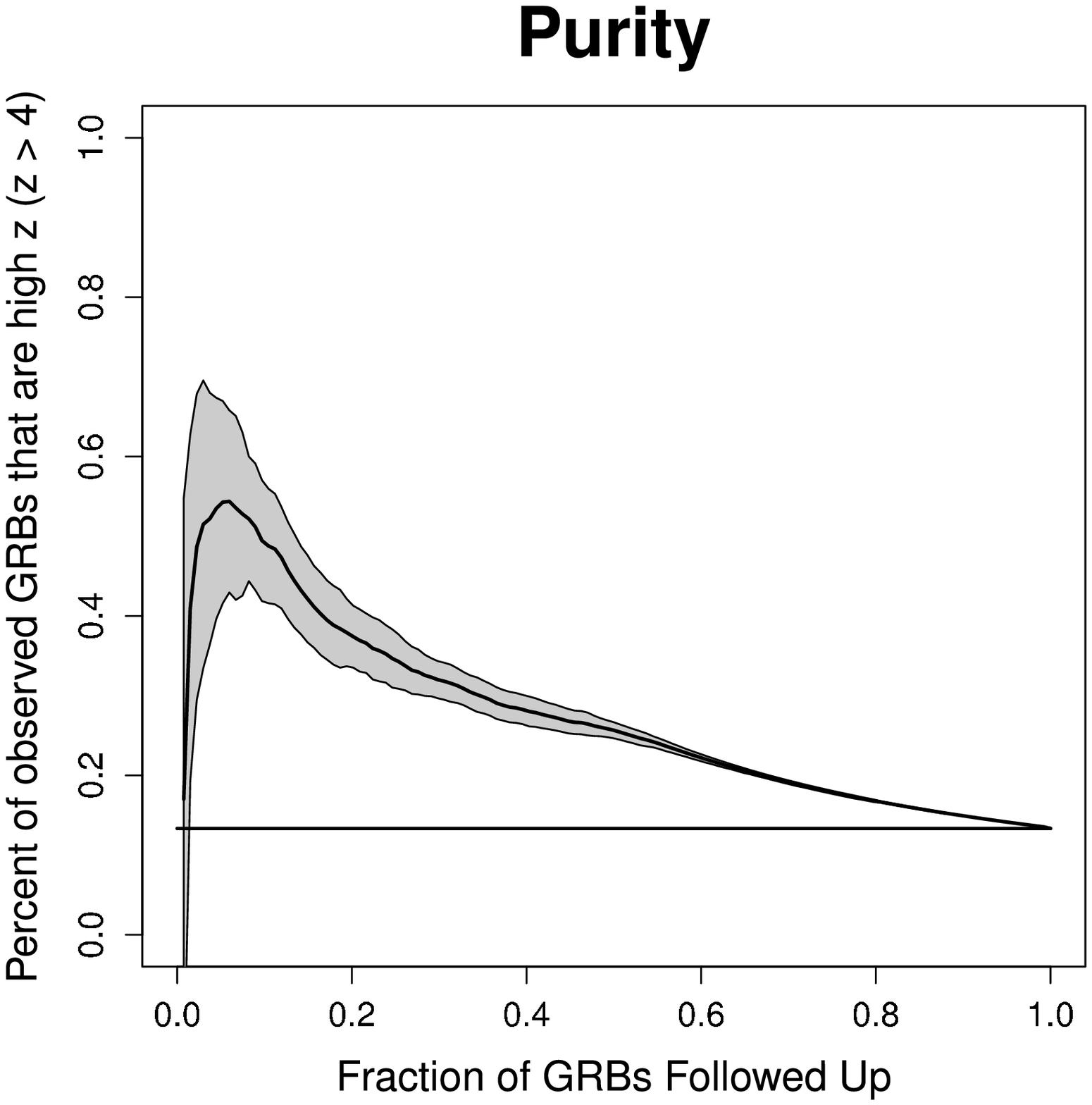}}
\  \caption[?????]
  {Efficiency ($N_{\rm high~observed}/N_{\rm total~high}$; left panel) and purity ($N_{\rm high~observed}/N_{\rm total~observed}$; right panel) versus fraction of GRBs followed up according to our decision criterion ($\mathcal{Q}$)with a high-$z$ cutoff of $z=4$. 18 bursts ($\sim13\%$ of our training set) are $z\ge4.0$.  The curve uncertainties shown are $1 \sigma$ standard deviations from the mean value across all seeds.
 }
\label{fig:efficiencypurity4}
\end{figure}

Taking into account the above issues of multiple feature set choices, the deleterious effect of useless features, and the performance with various weight choices to help with imbalance, we have developed a classifier which we believe to be robust and powerful. The full feature set utilized is shown in Table \ref{tab:features}, and the weight chosen is described in \S\ref{sec:weights}.  The final cross-validated estimates of $\widehat{\mathcal{Q}}$ for the training data are shown alongside the corresponding redshifts in Table \ref{tab:trainingredshifts}. By referencing a particular point on the $x$-axis of Figure \ref{fig:efficiencypurity4} (left panel) one can determine what fraction of high bursts can be detected for a particular amount of telescope follow-up time. For example, if we are able to follow up on 20\% of all GRBs detected by \textit{Swift}, then the bursts recommended for follow-up by our classifier will contain on average $56\% \pm 6\%$  of all GRBs with redshift greater than 4 that occur. Following-up on $\sim 40\%$ of all bursts will yield $84\% \pm 6\% $ of all GRBs with redshift greater than 4, and following-up on the top $50\%$ of candidates will result in nearly all of the high-$z$ events being observed ($96\% \pm 4\%$).

Purity is shown in the right panel of Figure \ref{fig:efficiencypurity4}, which describes how many of the followed-up bursts will actually be high-redshift. Following up on 20\% of all bursts would result in  $37\% \pm 4\%$ of the followed-up events being high-redshift, and $28\% \pm 2\%$ of followed up bursts would be high-redshift if $40\%$ of GRBs were followed-up on.

\begin{figure}[t!]
  \centerline{\plottwo{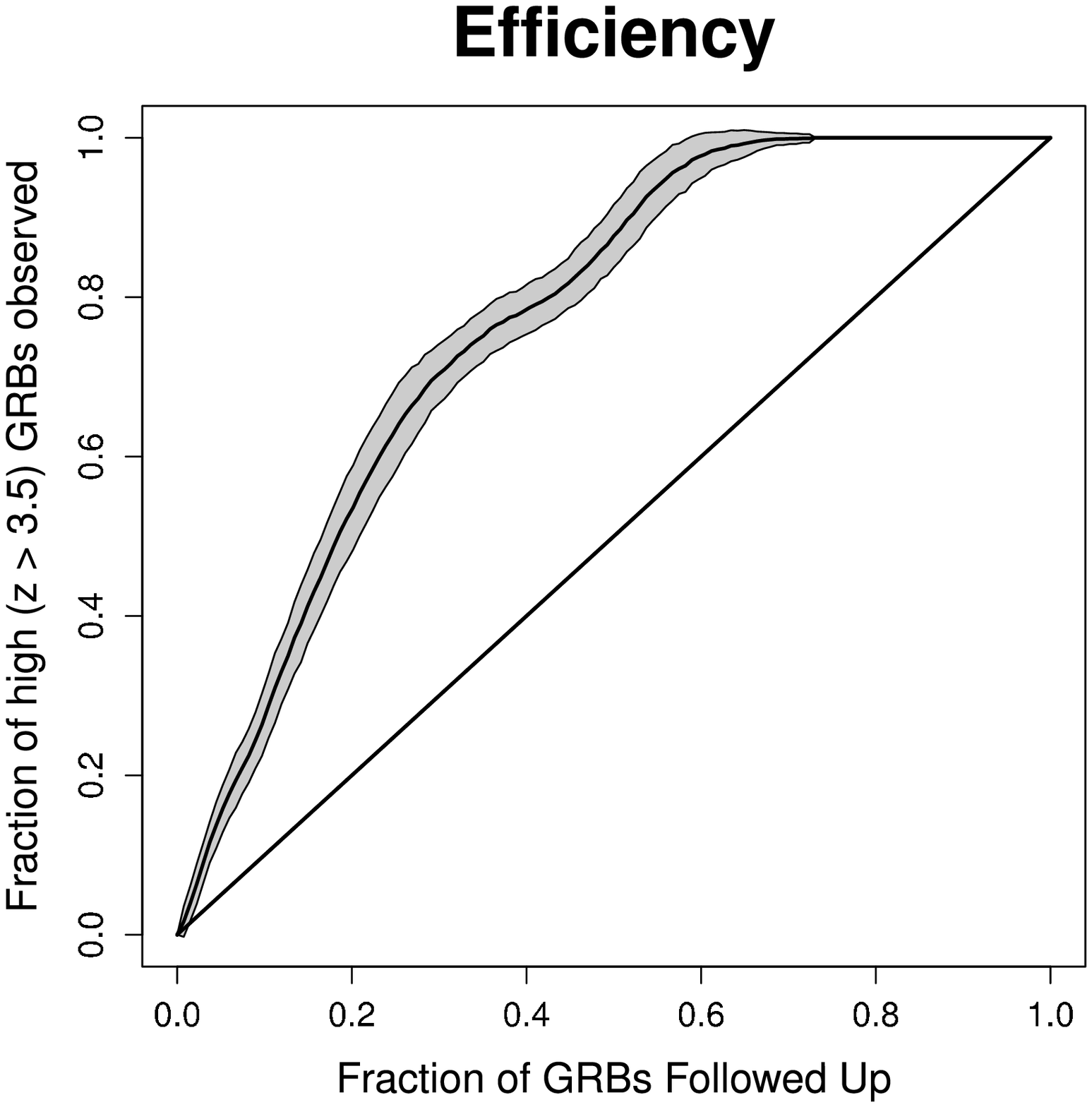}
					{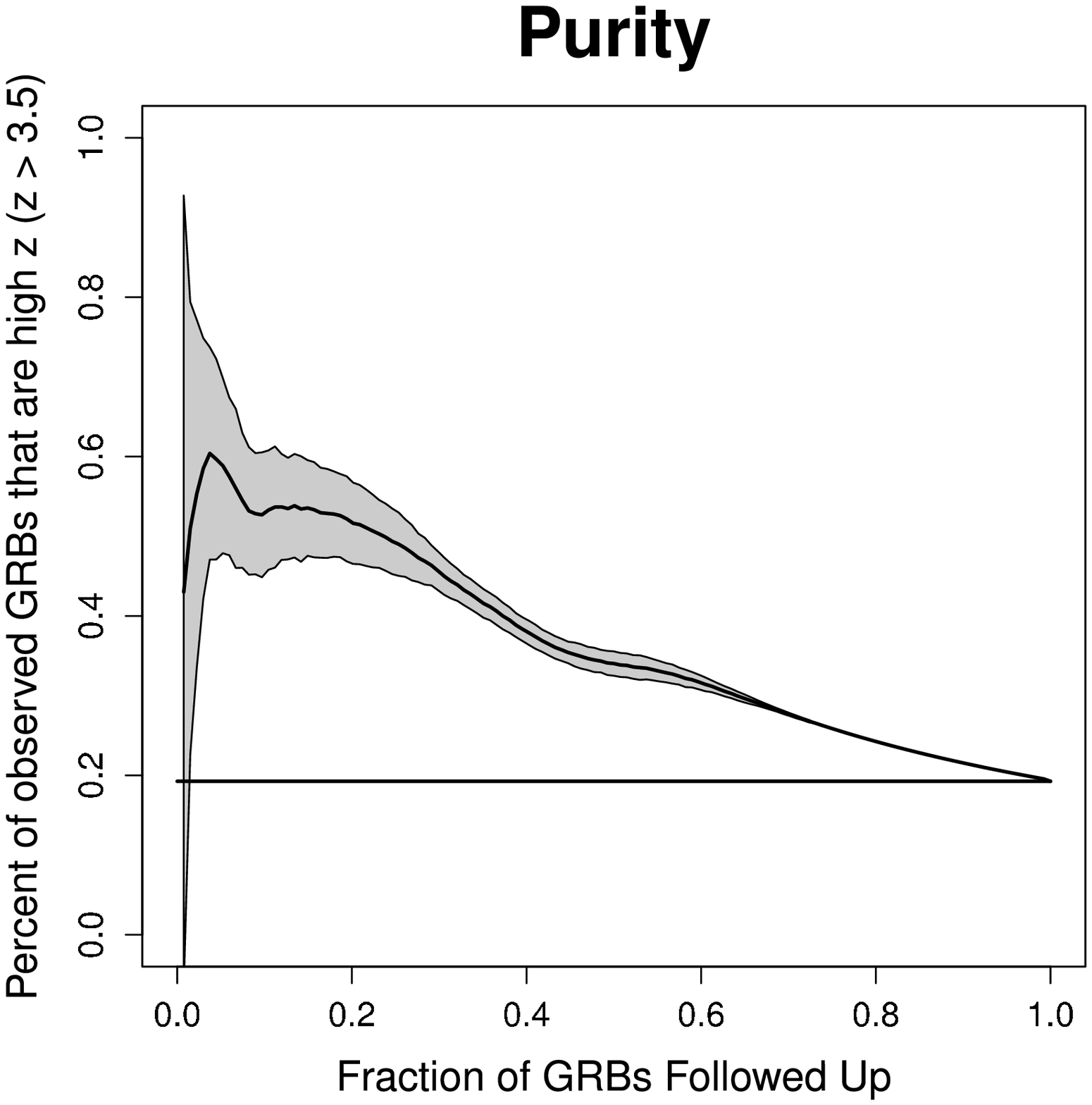}}
\  \caption[?????]
  {Efficiency ($N_{\rm high~observed}/N_{\rm total~high}$; left panel) and purity ($N_{\rm high~observed}/N_{\rm total~observed}$; right panel) versus fraction of GRBs followed up according to our decision criterion ($\mathcal{Q}$)with a high-$z$ cutoff of $z=3.5$. 26 bursts ($\sim19\%$ of our training set) are $z\ge3.5$.  The curve uncertainties shown are $1 \sigma$ standard deviations from the mean value across all seeds.
 }
\label{fig:efficiencypurity35}
\end{figure}

\begin{figure}[t!]
  \centerline{\plottwo{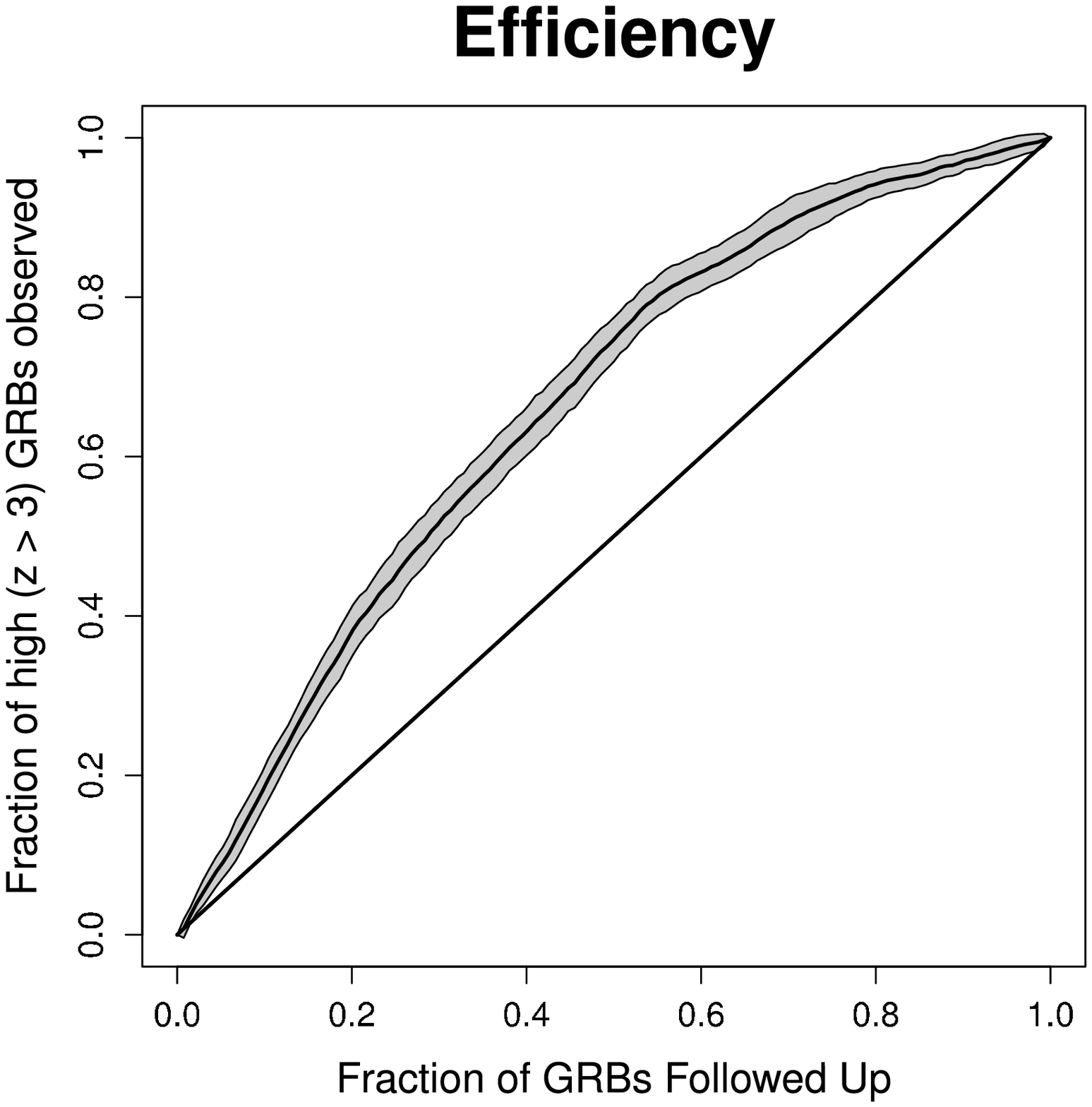}
					{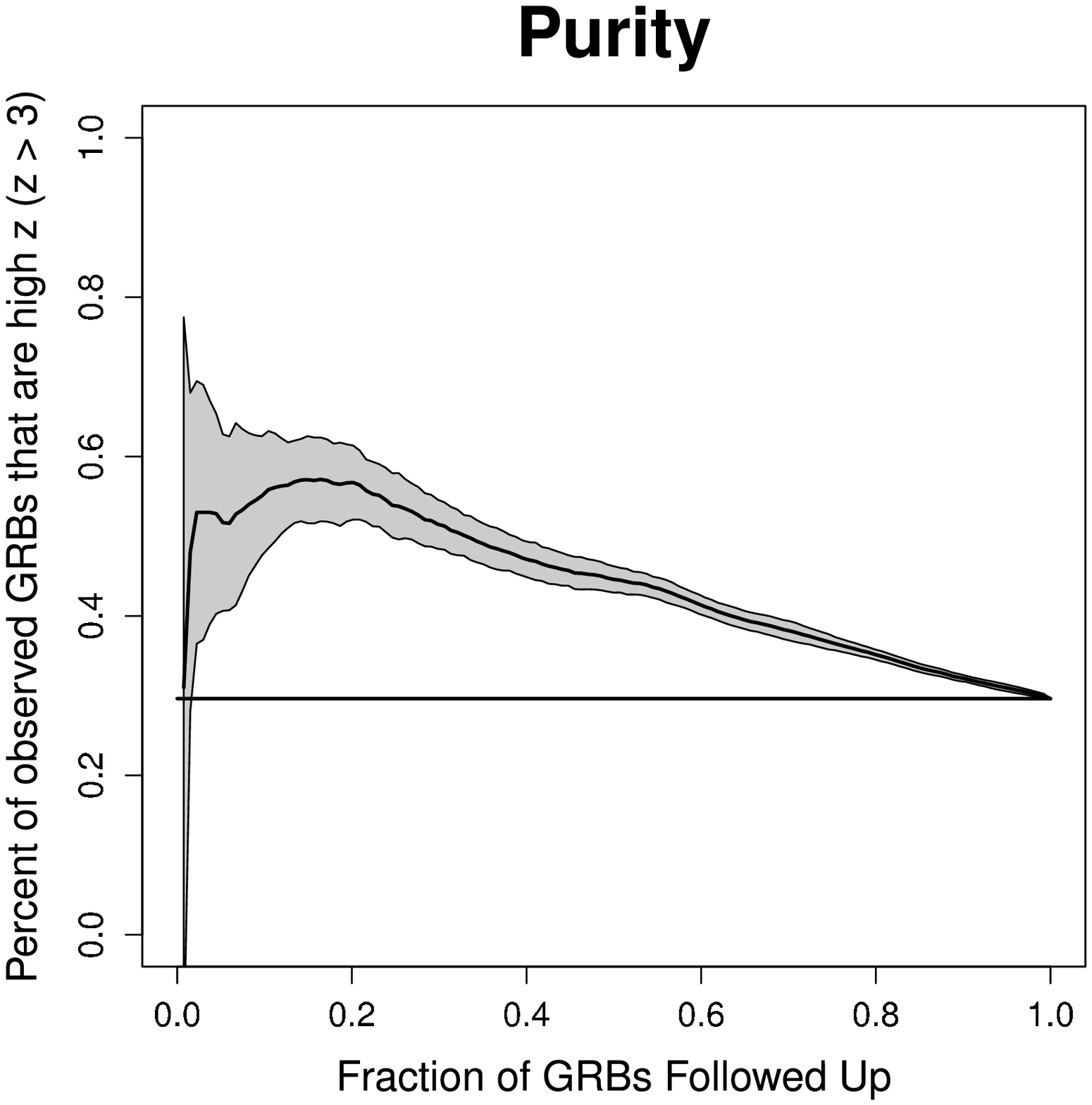}}
\  \caption[?????]
  {Efficiency ($N_{\rm high~observed}/N_{\rm total~high}$; left panel) and purity ($N_{\rm high~observed}/N_{\rm total~observed}$; right panel) versus fraction of GRBs followed up according to our decision criterion ($\mathcal{Q}$)with a high-$z$ cutoff of $z=3.0$. 40 bursts ($\sim30\%$ of our training set) are $z\ge3.0$.  The curve uncertainties shown are $1 \sigma$ standard deviations from the mean value across all seeds.
 }
\label{fig:efficiencypurity3}
\end{figure}

As the high/low class division of $z=4$ was relatively arbitrary, for completeness we also re-trained the classifier and calculated performance results using cutoff values of $z=3.5$ (Fig. \ref{fig:efficiencypurity35}) and $z=3$ (Fig. \ref{fig:efficiencypurity3}). Note that while the sample size of `high' events more than doubles by lowering the cutoff value to $z=3$, the resultant efficiency decreases significantly. We attribute this effect to a decrease in the predictive power of certain attributes at lower redshift. For instance, the $z > 3$ population has proportionally many more instances of UVOT detections in its `high-$z$' class than the $z > 4$ population, which reduces its effectiveness as a discriminating feature.


\subsection{Feature Importance}
\label{sec:importance}
\begin{figure}[t!]
  \centerline{\plottwo{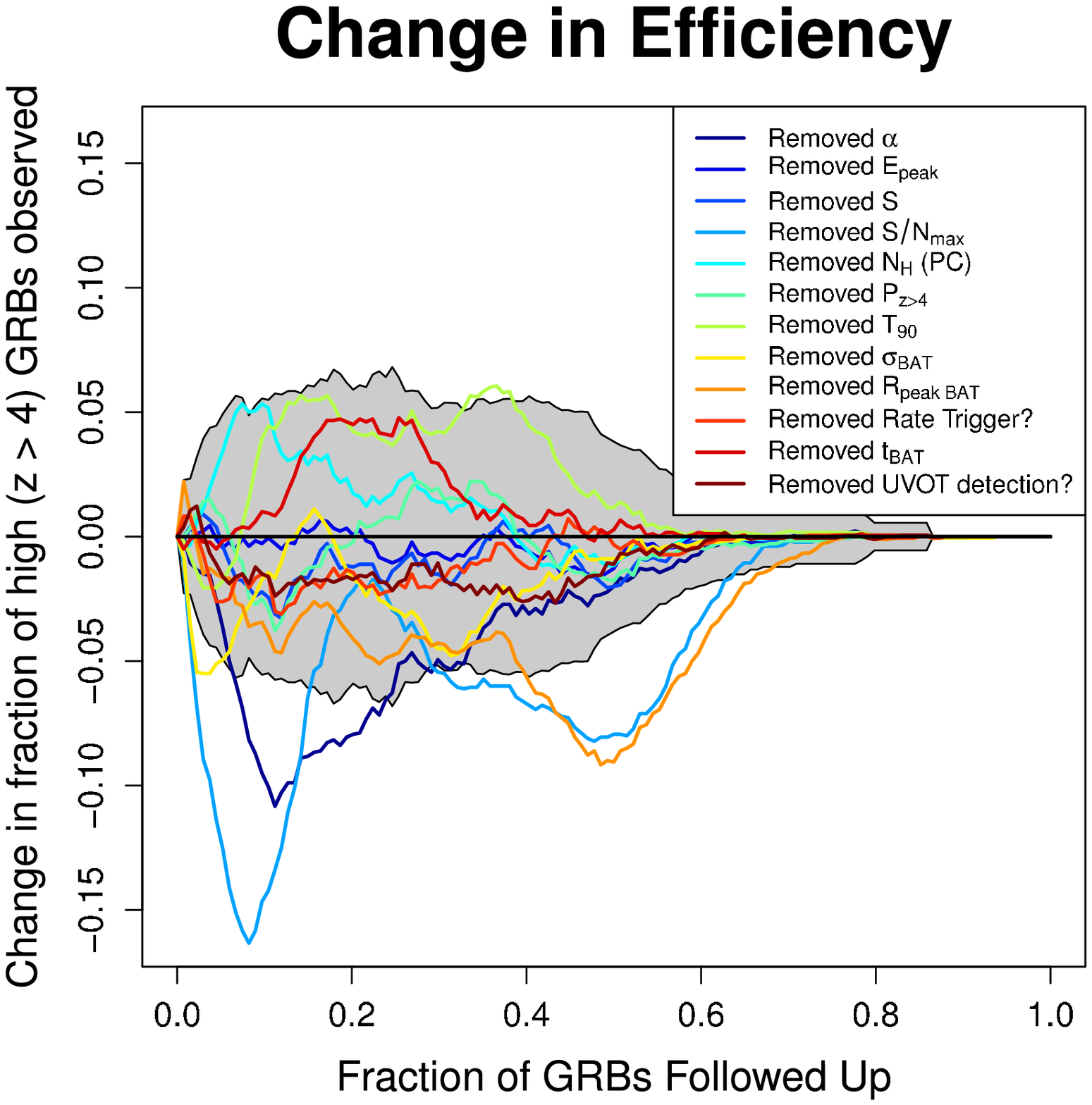}
			{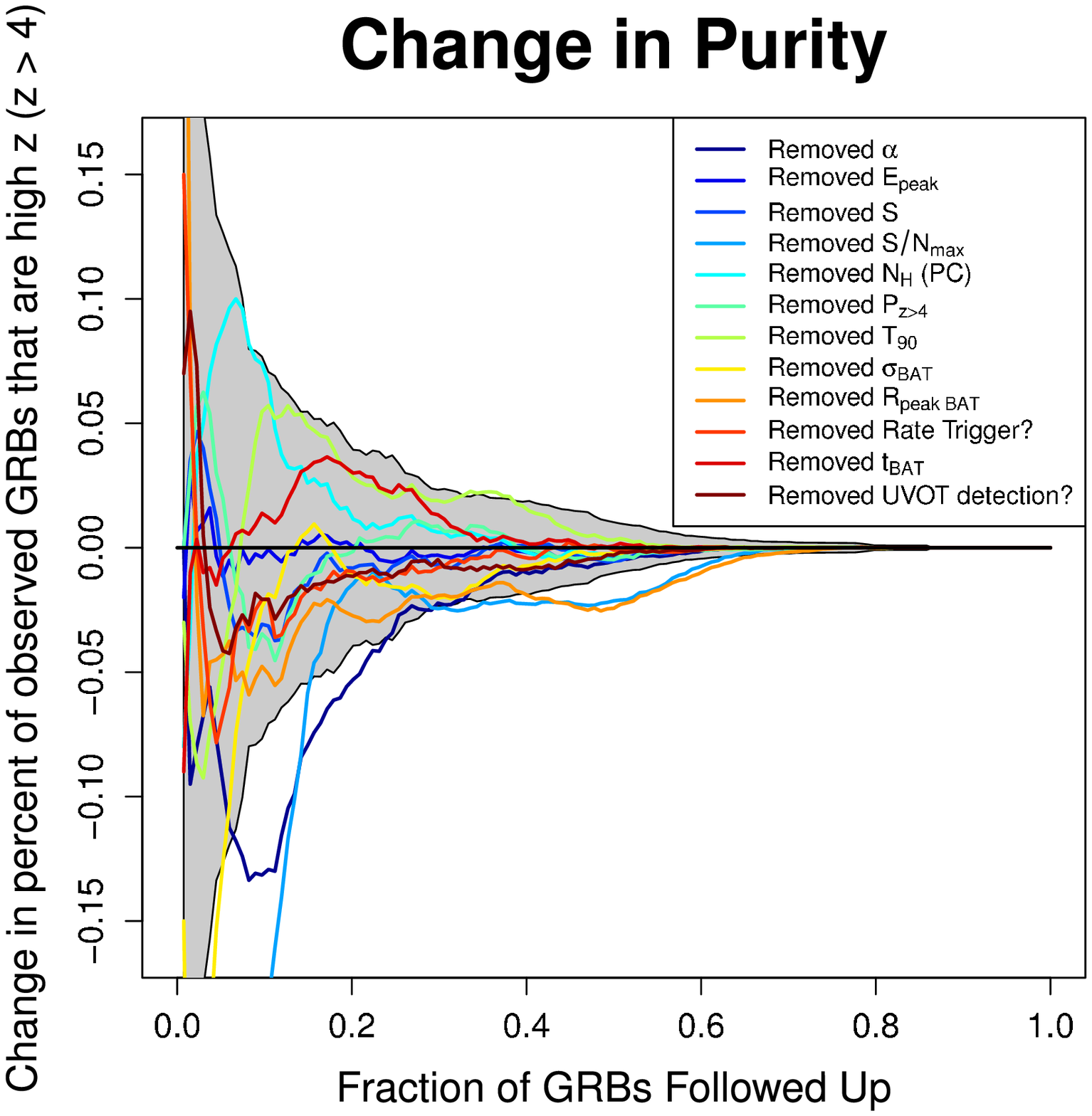}}
\  \caption[?????]
  {Change in efficiency (left) and purity (right) by removing individual features from the default feature set listed in Table \ref{tab:features}. The standard deviation from the mean value across all seeds for the default dataset is shown in grey.  The lack of degradation in performance by the removal of a feature does not necessarily imply that it has no predictive power, only that it may be redundant with other features.   Most of the features do not cause a significant change in performance once removed from the dataset.  However, the removal of a few of the individual features does cause a degradation in performance larger than what would be expected by random, implying that these features are both important and not completely redundant. Note that the relative change in both purity and efficiency are equal in both plots, as only the numerator of each metric is changing ($N_{\rm high~observed}$), but we show both values for consistency.
 }
\label{fig:featurecompare}
\end{figure}

There are several complications in identifying the relative importance of features in contributing to selecting high-$z$ candidates.  To an extent, simple scatter plots such as those in Figure \ref{fig:featuresvfeatures} can give an indication as to what features are best at separating the classes, but these fail to account for the complex interactions between features occurring within the RF classification.  The effects of removing features from the dataset and then re-constructing the classifier give another indication of feature importance, but fail to account for redundancy in the features; if two features have similar predictive properties, removing one will just cause the other to take its place.  Nevertheless, such an experiment can be illustrative, and the results are shown in Figure \ref{fig:featurecompare}.  In general, the removal of an individual feature does not cause a significant change in performance, and the small changes that do occur trend toward a degradation in the number of high-$z$ bursts identified, implying that few if any of the features in the dataset are useless. The features that cause the largest degradation in performance upon their removal are $\alpha, R_{peak,BAT},$ and $S/N_{\rm max},$ indicating that these features are both useful predictors and are not fully redundant with other features. Note that the slight improvement in performance from the removal of the temporal features $T_{90}$ and  $t_{BAT}$ is consistent with these values having little-to-no predictive power, in agreement with the recent findings of \citet{kocevski11a} showing a lack of time dilation signatures in GRB light curves.

\section{Discussion}
\label{sec:application}

\subsection{Calibration on GRBs with unknown redshifts}
\label{sec:unknownz}
A natural application of our methodology is to use it to predict the follow-up metric $\widehat{\mathcal{Q}}$ for the remaining majority of long-duration \Swift{}  GRBs with no known redshift, providing a list of the top candidates predicted to be high-$z$.  This application is precisely how \rategrbz{} could be used in practice on new events, albeit one-at-a-time rather than on many at once.  We caution that due to the natural selection effect of GRBs with measured redshifts having a higher likelihood of being brighter events, the bursts with unknown redshifts are likely to comprise a somewhat different redshift distribution than our training dataset.  The primary consequence of this is the interpretation of the user-desired follow-up fraction $\mathcal{Q}$ and the prioritization parameter $\widehat{\mathcal{Q}}$.  In principle, the classifier was calibrated such that, over time, a fraction $\mathcal{Q}$ of new events will have affirmative follow-up recommendations (that is, events such that $\widehat{\mathcal{Q}} \le \mathcal{Q}$).  However, this will not necessarily be the case if the full redshift distribution of GRBs makes up a different population than our training data.

\begin{figure}[t!]
 \centerline{\plottwo{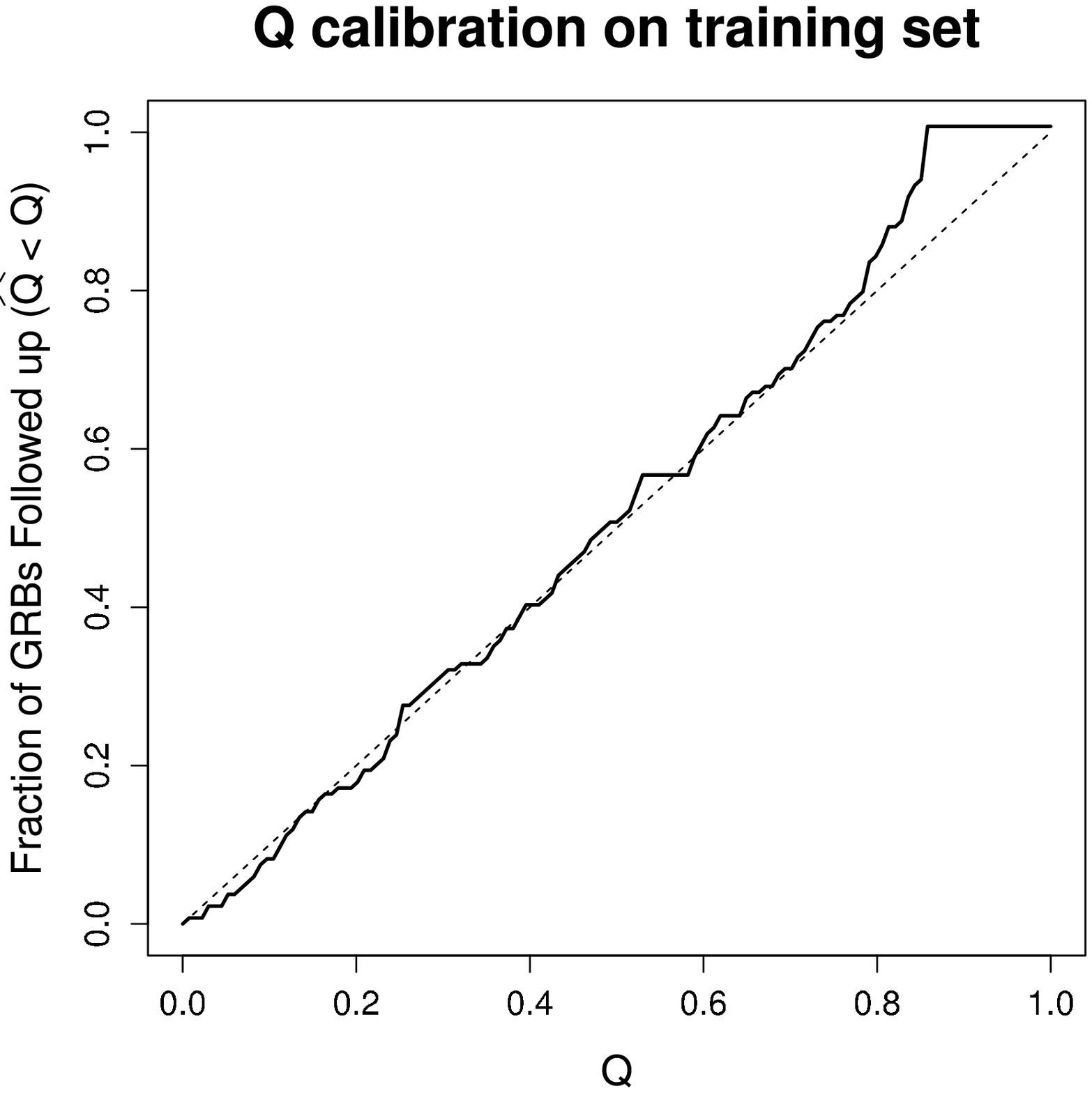}{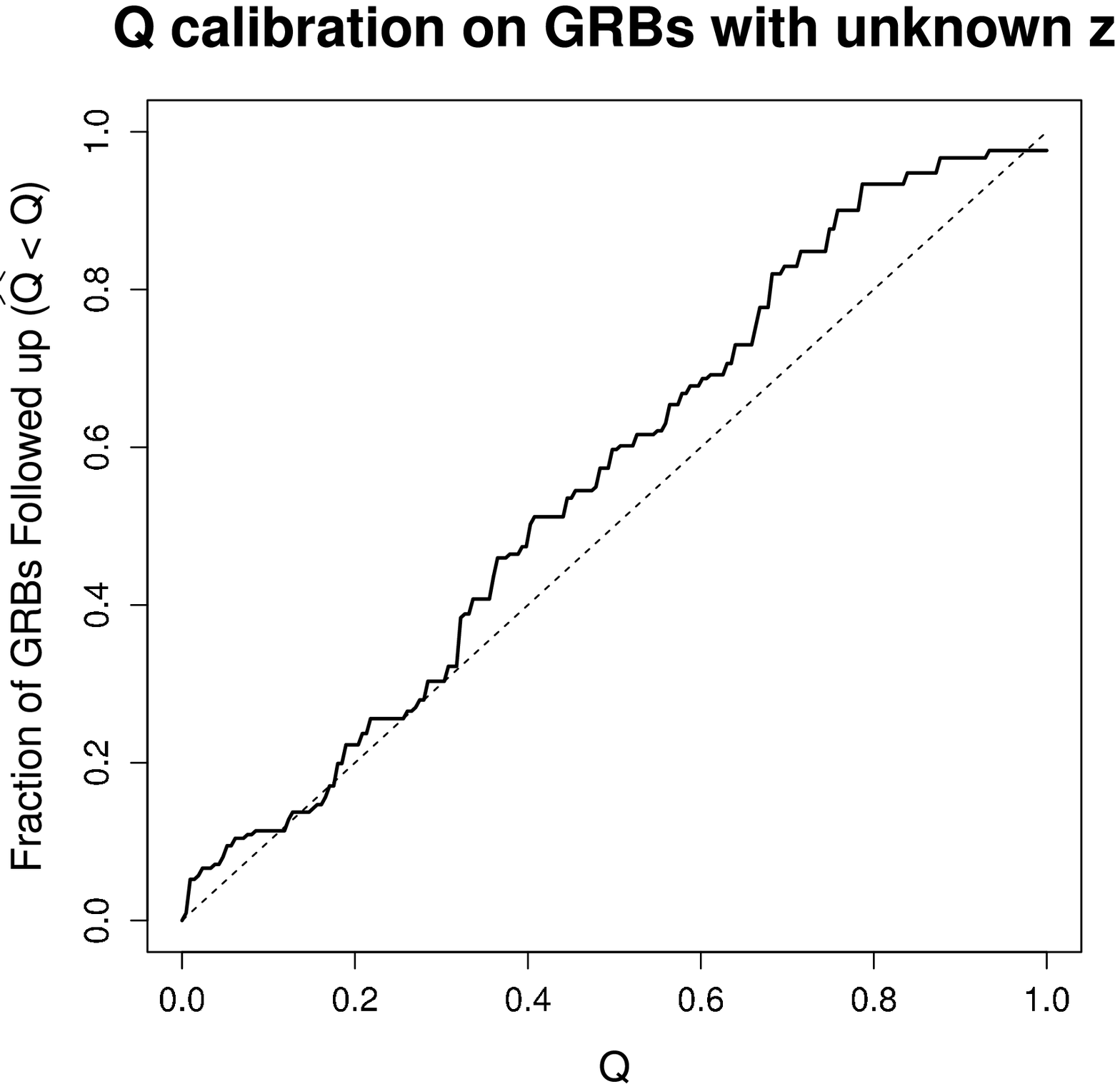}}
\  \caption[?????]
  {Here we quantify the calibration of $\mathcal{Q}$; namely, how well does the user-desired follow-up fraction $\mathcal{Q}$ correspond to the actual number of bursts recommended to be followed up by the algorithm ($\widehat{\mathcal{Q}} \le \mathcal{Q}$). The left figure shows the self-calibration of the cross-validated training set, which aligns as expected.  The right plot shows the calibration on the test set is good, especially at low $\mathcal{Q}$.  At larger $\mathcal{Q}$, there is a slight departure from the diagonal, implying a follow-up recommendation of more events than expected at these values. This can be attributed to the differing populations between the training set (with measured redshifts) and test set (with unknown redshifts), as illustrated in Figure \ref{fig:featuresvfeatures}.  This slight discrepancy is not surprising, as low brightness events without UVOT detections are naturally more difficult to obtain redshifts for.}
\label{fig:popcompare_unknownz}
\end{figure}

To test this, we calculated $\widehat{\mathcal{Q}}$ for each of the remaining 212 GRBs with unknown redshift that met our culling criteria outlined in \S \ref{sec:obs}.  From this we could calculate the fraction of GRBs followed up ($\widehat{\mathcal{Q}} \le \mathcal{Q}$) for each cutoff value of $\mathcal{Q}$.  The results of this test are shown in Figure \ref{fig:popcompare_unknownz}.  For the chosen weight of 10 (see \S\ref{sec:weights}), the $\mathcal{Q}$-values are well calibrated with the final follow-up recommendations. The resultant $\widehat{\mathcal{Q}}$ priorities are listed in Table \ref{tab:unknown}. These values can be interpreted as a ranking of which of these past events without secure redshift determinations are most likely to be at high-redshift.

\begin{landscape}
\thispagestyle{empty}
            \begin{deluxetable}{llllllllllllll}
        \tabletypesize{\scriptsize}
        \singlespace
        \tablewidth{0pt}
        \tablecaption{Test Data}
        \tablehead{\colhead{GRB} &
\colhead{$\widehat{\mathcal{Q}}$} &
\colhead{$\alpha$} &
\colhead{$E_{peak}$} &
\colhead{$S$} &
\colhead{$S/N_{max}$} &
\colhead{$N_{H,pc}$} &
\colhead{$T_{90}$} &
\colhead{$\sigma_{BAT}$} &
\colhead{$R_{peak,BAT}$} &
\colhead{Rate} &
\colhead{$t_{BAT}$} &
\colhead{UVOT} &
\colhead{$P_{z>4}$} \\
\colhead{} &
\colhead{} &
\colhead{} &
\colhead{(keV)} &
\colhead{(erg/cm$^2$)} &
\colhead{} &
\colhead{$(10^{22}$ cm$^{-2})$} &
\colhead{(s)} &
\colhead{} &
\colhead{(ct/s)} &
\colhead{trigger} &
\colhead{(s)} &
\colhead{detect} &
\colhead{} }
\startdata
050215A	&	3.19e-01	&	-1.29e+00	&	4.14e+02	&	1.34e-06	&	1.02e+01	&	?	&	6.65e+01	&	9.00e+00	&	6.94e+02	&	yes	&	8.19e+00	&	no	&	9.81e-02 \\ 
050215B	&	1.78e-01	&	?	&	3.01e+01	&	2.86e-07	&	1.44e+01	&	5.70e-02	&	8.50e+00	&	8.00e+00	&	3.00e+02	&	yes	&	2.05e+00	&	no	&	1.06e-01 \\ 
050219A	&	4.22e-01	&	1.87e-02	&	1.00e+02	&	4.91e-06	&	5.08e+01	&	9.10e-02	&	2.50e+01	&	8.00e+00	&	1.93e+02	&	yes	&	1.02e+00	&	no	&	1.12e-01 \\ 
050219B	&	7.33e-01	&	-8.94e-01	&	1.12e+02	&	1.94e-05	&	7.19e+01	&	8.80e-02	&	2.09e+01	&	1.70e+01	&	4.09e+02	&	yes	&	1.02e+00	&	no	&	2.73e-02 \\ 
050326	&	7.04e-01	&	-1.04e+00	&	3.41e+02	&	1.70e-05	&	1.33e+02	&	3.80e-02	&	3.02e+01	&	2.10e+01	&	1.84e+04	&	yes	&	5.12e-01	&	no	&	5.67e-02 
\enddata
\tablecomments{Table \ref{tab:unknown} is published in its entirety in the electronic edition of The Astrophysical Journal. A portion is shown here for guidance regarding its form and content.}
\label{tab:unknown}
\end{deluxetable}
\end{landscape}

\subsection{Validation Set: Application to Recent GRBs}

Since the cutoff date in our training set (June 21, 2010) until Sept. 1, 2011, there have been 15 long duration \Swift{} GRBs with reliable redshifts from which we constructed an independent validation set to test our method\footnote{One of the bursts with a measured redshift, GRB 110328A, had very unusual properties and was determined to be a potential Tidal Disruption Event \citep{bloom11,levan11}, and was thus also excluded from the validation set.}.
The feature values for these GRBs are presented in Table \ref{tab:validation}. 
While none of these events were over our high-redshift cutoff value of $z=4$, it is still possible, though challenging, to use low-$z$ events (either by direct redshift measurement or by the identification of a coincident blue host galaxy) as a consistency test.   We would expect that the purity at a given $\mathcal{Q}$ would be lower than the fraction of recommended follow-up events ($\widehat{\mathcal{Q}} \le \mathcal{Q}$) \emph{without} a secure low-$z$ determination. For instance, $\mathcal{Q} = 0.2$ has a purity of $37\% \pm 4\%$, so no more than $\sim63\%$ of events with $\widehat{\mathcal{Q}} < 0.2$ should be definitively low-redshift.

The validation GRBs were run through the \rategrbz{} classifier, and their resultant $\widehat{\mathcal{Q}}$ values are shown in Table \ref{tab:validationredshifts} along with their corresponding redshifts. The smallest $\widehat{\mathcal{Q}}$ value of these events is $\sim 0.3$, meaning that none of these events would have been recommended for high-$z$ follow-up for anyone wishing to observe fewer than 30\% of events.
While these values are certainly consistent with our expected purity, it is not particularly constraining, as it would have been very unlikely for this almost-random selection of GRBs to violate this constraint by chance alone, even if the classifier had no predictive power. 

A more constraining test is the identification of high-$z$ events with high $\widehat{\mathcal{Q}}$ for comparison with the expected efficiency.  
Two events not included in our training set have had recent high-$z$ identifications: GRB 090429B with strong photometric evidence for being $z \simeq 9.4$ \citep{Cucchiara11}, and the spectroscopic identification of GRB 111008A at $z=4.99$ \citep{levan11a,wiersema11a}. The former has a $\widehat{\mathcal{Q}}$ value of $\sim0.185$, consistent with the expected efficiency. However, GRB 111008A has a $\widehat{\mathcal{Q}}$ of $\sim0.637$, a value above which we would have expected to find no more than $~1\%$ of high-$z$ events.  This outlier seems likely due to the extreme brightness of the event (among the brightest $\sim10\%$ of \Swift{} bursts in the observer frame, and top $\sim3\%$ in the rest frame). Indeed, compared to all 18 high-$z$ events in the training set, GRB 111008A has the most extreme values towards the `wrong' end of three of the highly important features identified in \S\ref{sec:importance} ($\alpha, P_{z>4}$, and $R_{peak,BAT}$) and also has the fourth largest $S/N_{max}$.
In later iterations of \rategrbz, this event (and all new GRBs with secure redshifts) will be added to the training data to re-generate the classifier and further improve its robustness against such outliers.

\begin{landscape}
\thispagestyle{empty}
            \begin{deluxetable}{llllllllllllll}
        \tabletypesize{\scriptsize}
        \singlespace
        \tablewidth{0pt}
        \tablecaption{Validation Data}
        \tablehead{\colhead{GRB} &
\colhead{$\widehat{\mathcal{Q}}$} &
\colhead{$\alpha$} &
\colhead{$E_{peak}$} &
\colhead{$S$} &
\colhead{$S/N_{max}$} &
\colhead{$N_{H,pc}$} &
\colhead{$T_{90}$} &
\colhead{$\sigma_{BAT}$} &
\colhead{$R_{peak,BAT}$} &
\colhead{Rate} &
\colhead{$t_{BAT}$} &
\colhead{UVOT} &
\colhead{$P_{z>4}$} \\
\colhead{} &
\colhead{} &
\colhead{} &
\colhead{(keV)} &
\colhead{(erg/cm$^2$)} &
\colhead{} &
\colhead{$(10^{22}$ cm$^{-2})$} &
\colhead{(s)} &
\colhead{} &
\colhead{(ct/s)} &
\colhead{trigger} &
\colhead{(s)} &
\colhead{detect} &
\colhead{} }
\startdata
100728B	&	6.07e-01	&	-1.64e+00	&	8.19e+01	&	2.54e-06	&	2.06e+01	&	3.90e-02	&	1.15e+01	&	9.07e+00	&	1.47e+02	&	yes	&	1.02e+00	&	yes	&	1.01e-01 \\ 
100814A	&	6.81e-01	&	-1.11e+00	&	1.35e+02	&	9.33e-06	&	9.80e+01	&	?	&	1.77e+02	&	1.91e+01	&	8.34e+02	&	yes	&	1.02e+00	&	yes	&	1.80e-01 \\ 
100816A	&	9.33e-01	&	-5.71e-01	&	1.42e+02	&	2.71e-06	&	5.80e+01	&	1.13e-01	&	2.50e+00	&	2.29e+01	&	1.42e+03	&	yes	&	1.02e+00	&	yes	&	5.55e-02 \\ 
100901A	&	4.00e-01	&	-1.55e+00	&	1.28e+02	&	3.41e-06	&	1.78e+01	&	4.00e-02	&	4.59e+02	&	7.70e+00	&	4.50e+02	&	yes	&	8.19e+00	&	yes	&	2.25e-01 \\ 
100906A	&	1.00e+00	&	-1.66e+00	&	1.57e+02	&	1.37e-05	&	1.36e+02	&	?	&	1.17e+02	&	1.05e+01	&	1.91e+02	&	yes	&	5.12e-01	&	yes	&	7.39e-02 \\ 
101219B	&	6.30e-01	&	-1.89e+00	&	4.97e+01	&	3.75e-06	&	1.00e+01	&	-8.00e-03	&	4.18e+01	&	7.63e+00	&	8.44e+02	&	no	&	6.40e+01	&	yes	&	1.07e-01 \\ 
110205A	&	3.19e-01	&	-1.39e+00	&	9.75e+01	&	1.98e-05	&	1.50e+02	&	1.10e-02	&	2.77e+02	&	1.00e+01	&	1.48e+03	&	no	&	6.40e+01	&	yes	&	1.45e-01 \\ 
110213A	&	9.33e-01	&	-1.82e+00	&	6.70e+01	&	8.77e-06	&	3.10e+01	&	4.00e-02	&	4.31e+01	&	1.21e+01	&	2.05e+02	&	yes	&	1.02e+00	&	yes	&	5.32e-02 \\ 
110422A	&	1.00e+00	&	-6.23e-01	&	1.11e+02	&	5.17e-05	&	2.10e+02	&	1.58e-01	&	2.67e+01	&	7.19e+00	&	8.20e+01	&	yes	&	1.28e-01	&	yes	&	2.49e-02 \\ 
110503A	&	9.33e-01	&	-8.18e-01	&	1.42e+02	&	1.43e-05	&	6.27e+01	&	2.60e-02	&	9.31e+00	&	2.04e+01	&	1.26e+03	&	yes	&	1.02e+00	&	yes	&	1.89e-02 \\ 
110715A	&	9.33e-01	&	-1.06e+00	&	8.94e+01	&	1.40e-05	&	2.02e+02	&	1.64e-01	&	1.31e+01	&	1.19e+01	&	1.47e+02	&	yes	&	1.28e-01	&	yes	&	9.70e-03 \\ 
110726A	&	5.04e-01	&	-2.97e-01	&	4.27e+01	&	2.07e-07	&	1.51e+01	&	-4.90e-02	&	5.40e+00	&	8.60e+00	&	2.24e+02	&	yes	&	1.02e+00	&	yes	&	1.14e-01 \\ 
110731A	&	1.00e+00	&	-1.19e+00	&	4.06e+02	&	1.25e-05	&	1.30e+02	&	7.20e-02	&	4.66e+01	&	2.46e+01	&	2.32e+03	&	yes	&	1.02e+00	&	yes	&	5.09e-02 \\ 
110801A	&	9.33e-01	&	-1.84e+00	&	6.07e+01	&	6.85e-06	&	3.56e+01	&	2.90e-02	&	4.00e+02	&	7.83e+00	&	3.50e+02	&	yes	&	4.10e+00	&	yes	&	1.98e-01 \\ 
110808A	&	5.56e-01	&	?	&	2.59e+01	&	4.27e-07	&	1.01e+01	&	2.17e-01	&	3.94e+01	&	7.19e+00	&	4.26e+02	&	yes	&	8.19e+00	&	yes	&	1.06e-01
\enddata
\label{tab:validation}
\end{deluxetable}
\end{landscape}

\begin{deluxetable}{llll}
        \tabletypesize{\scriptsize}
        \singlespace
        \tablewidth{0pt}
        \tablecaption{Validation Redshifts and Predictions}
        \tablehead{\colhead{GRB} &
\colhead{$\widehat{\mathcal{Q}}$} &
\colhead{$z$} &
\colhead{References} }
\startdata
100728B	&	5.63e-01	&	2.106	&	\citealt{flores10a} \\ 
100814A	&	7.04e-01	&	1.44	&	\citealt{omeara10a} \\ 
100816A	&	9.33e-01	&	0.8035	&	\citealt{tanvir10a,tanvir10b} \\ 
100901A	&	4.22e-01	&	1.408	&	\citealt{chornock10c} \\ 
100906A	&	9.19e-01	&	1.727	&	\citealt{tanvir10c} \\ 
101219B	&	6.07e-01	&	0.5519	&	\citealt{postigo11a} \\ 
110205A	&	3.19e-01	&	2.22	&	\citealt{cenko11a} \\ 
110213A	&	8.89e-01	&	1.46	&	\citealt{milne11a} \\ 
110422A	&	9.33e-01	&	1.770	&	\citealt{malesani11a,postigo11b} \\ 
110503A	&	9.33e-01	&	1.613	&	\citealt{postigo11c} \\ 
110715A	&	8.89e-01	&	0.82	&	\citealt{piranomonte11a} \\ 
110726A	&	5.56e-01	&	1.036	&	\citealt{cucchiara11a} \\ 
110731A	&	9.33e-01	&	2.83	&	\citealt{tanvir11a} \\ 
110801A	&	8.67e-01	&	1.858	&	\citealt{lavers11a} \\ 
110808A	&	5.63e-01	&	1.348	&	\citealt{postigo11d} 
\enddata
\label{tab:validationredshifts}
\end{deluxetable}

\subsection{Comparison to Previous Efforts}

Extracting indications of redshift from promptly available information has been a continuing goal of GRB studies since their cosmological origins were discovered nearly 15 years ago.  Several potential luminosity indicators were pursued with the optimistic goal of using GRBs as standard candles for cosmological studies.  The efficacy of individual indicators toward this goal proved to be limited, and a physical origin of the relations has been contested, with authors attributing them instead to detector thresholding or other selection effects \citep{butler07b,butler09,butler10,shahmoradi11}.  While these studies have ruled out the majority of such relations as intrinsic to GRBs themselves, prompt properties can still be used as redshift indicators if the systematics are properly accounted for. 

Several recent studies have attempted to use combinations of features to determine ``pseudo-redshifts'' for GRBs. In an extension of work by \citet{schaefer07}, \citet{xiao09,xiao11} used a combination of six purported luminosity relations. Further, \citet{koen09, koen10} has explored linear regression as a tool for predicting GRB redshifts using the dataset from \citet{schaefer07}. As data derived from multiple satellites were used, these studies are particularly vulnerable to the detector selection effects mentioned above.  

Some works avoided the complications of regression and instead focused upon the simple selection of high-$z$ candidates for follow-up purposes. \citet{campana07} utilized a sample of \Swift-only bursts (thus avoiding detector effect biases) and used hard cuts on three features ($T_{90}$, lack of UVOT detection, and high-galactic latitude) for high-$z$ candidate selection.  \citet{salvaterra07} extended upon this work with the additional feature of peak photon flux.

Several issues prevent a direct comparison among the various methods of the effectiveness at separating high-$z$ events. These include the usage of different features from each study, which is complicated by the lack of uniformity of features being created for each.  Further, the techniques above strictly constrain the manner in which each feature influences the output, whereas our method is fully non-parametric and therefore more flexible. However, the largest concern is accurate reporting of predictive performance. In particular, we caution against the circular practice of measuring the performance of methods by applying them to the same events from which the luminosity relations were formed.  In order to prevent over-estimating the accuracy of a predictive model, one needs to test on data independent from the training set, such as with cross-validation.

Finally, the \rate{} method differs from previous efforts in that it casts the problem as one of optimal resource allocation under limited follow-up time.  Prior techniques are not explicitly calibrated to suit this purpose. Direct classification methods will either under or over-utilize available resources. Past regression or ``pseudo-$z$'' methods are not explicitly calibrated to a particular follow-up decision (i.e., at what ``pseudo-$z$'' does one decide to follow up?), though it would be possible in principle to correct for this using a transformation which ensures that the desired follow-up fraction corresponds to the actual fraction of bursts followed up (e.g., Figure \ref{fig:popcompare_unknownz}). In contrast, the \rate{} technique is by design applicable to any available resource reserves, and is generally extendable to any transient follow-up prioritization problem.

\section{Conclusions}
\label{sec:conclusions}

In this paper, we presented the \rategrbz{}  method for allocating follow-up telescope resources to high-redshift GRB candidates using Random Forest classification on early-time \Swift{}  metrics. The \rate{}  method is generalizable to any prioritization problem that can be parameterized as ``observe'' or ``don't observe'', and accommodates statistical challenges such as small datasets, imbalanced classes, and missing feature values. The issue of resource allocation is becoming increasingly important in the era of data-driven transient surveys such as PTF, Pan-STARRS, and LSST which provide extremely high discovery rates without a significant increase in follow-up resources. With enough training instances of any object of interest for a given transient survey, the \rate{} method can be applied to prioritize follow-up of future high-priority candidates.

In the \rategrbz{}  application, our robust, cross-validated performance metrics indicate that by observing just 20\% of bursts, one can capture $56\% \pm 6\%$ of $z > 4$ events with a sample purity of $37\% \pm 4\%$.  Further,  following up on half of all events will yield nearly all ($96\% \pm 4\%$) of the high-$z$ events.  The method provides a simple decision point for each new event: if the prioritization value $\widehat{\mathcal{Q}}$ is smaller than the percent of events a user wishes to allocate resources to, then follow-up is recommended. These rapid predictions, combined with the more traditional photometric dropout technique from simultaneous multi-filter NIR observatories (such as PAIRITEL, GROND, and the upcoming RATIR), offer a robust tool in more efficiently informing GRB follow-up decisions.
To facilitate the dissemination of high-redshift GRB predictions to the community, we have set up a website (\url{http://rate.grbz.info}) with $\widehat{\mathcal{Q}}$ values for past bursts, and an RSS feed (\url{http://rate.grbz.info/rss.xml}) to provide real-time results from our classifier on new events.




\acknowledgments{
This work was sponsored by an NSF-CDI grant (award \#0941742) ``Real-time Classification of Massive Time-series Data Streams'' (PI: Bloom). This publication has made use of data obtained from the \Swift{}  interface
of the High-Energy Astrophysics Archive (HEASARC), provided by
NASA's Goddard Space Flight Center. A. Morgan gratefully acknowledges support from an NSF Graduate Research Fellowship. T. Broderick was funded by a National Science Foundation Graduate Research Fellowship.  We are extremely grateful to the \Swift{} team for the rapid public dissemination, calibration, and analysis of the \Swift{} data. We thank Daniel Perley for useful conversations and for the creation and maintenance of the very useful http://grbox.net website. We also thank Scott Barthelmy for his invaluable efforts in creating and maintaining the GCN system.}

{\it Facilities:}
\facility{Swift (BAT, XRT, UVOT)}


\end{document}